\documentclass[pdflatex,sn-vancouver-ay]{sn-jnl}
\usepackage{graphicx}%
\usepackage{multirow}%
\usepackage{amsmath,amssymb,amsfonts}%
\usepackage{amsthm}%
\usepackage{mathrsfs}%
\usepackage[title]{appendix}%
\usepackage{xcolor}%
\usepackage{textcomp}%
\usepackage{manyfoot}%
\usepackage{booktabs}%
\usepackage{algorithm}%
\usepackage{algorithmicx}%
\usepackage{algpseudocode}%
\usepackage{listings}%
\usepackage{bm}
\usepackage{setspace}
\usepackage{xurl}
\usepackage{placeins}

\begin{document}

%TC:ignore
\title[]{Quantifying internal variability in large-ensemble climate data with the Wasserstein distance}

\author*[1]{\fnm{Yuki} \sur{Yasuda}}\email{yuki.yasuda@jamstec.go.jp}
\author[1]{\fnm{Shoichiro} \sur{Kido}}\email{skido@jamstec.go.jp}
\affil[1]{\orgdiv{Research Institute for Earth and Information Sciences}, \orgname{Japan Agency for Marine-Earth Science and Technology}, \orgaddress{\street{3173-25 Showa-machi, Kanazawa-ku}, \city{Yokohama}, \postcode{2360001}, \state{Kanagawa}, \country{Japan}}}
%TC:endignore

\abstract{
    Estimating forced and internal variability in climate data is fundamental to detecting the climate change signal and assessing uncertainty in climate projections.
    We propose a metric that quantifies the relative magnitude of internal variability in single-model initial-condition large ensembles (SMILEs).
    We measure forced and internal variability by two 1-Wasserstein distances, a form of optimal transport cost.
    The proposed metric is the distance for internal variability divided by the sum of the two.
    Computing this ratio requires only sorting the samples and differencing the resulting quantiles, with no additional parameters.
    The metric reflects the entire distribution shape and applies to non-Gaussian variables, because the 1-Wasserstein distance measures the difference between any two probability distributions.
    We validate the metric with synthetic climate data from Gaussian, uniform, and lognormal distributions.
    Unlike the two existing metrics, the proposed metric gives stable estimates in the presence of outliers irrespective of the distribution shape, provided that the ensemble has about 40 members or more.
    We then apply the proposed and existing metrics to the 2~m air temperature and total precipitation of the Community Earth System Model Large Ensemble (CESM-LE) under two different forcing scenarios.
    All three metrics agree in the spatial pattern of the difference between the scenarios but not in amplitude.
    The amplitude of the difference is largest for the proposed metric, consistent with the synthetic-data result that this metric changes most rapidly when internal variability dominates.
    The 1-Wasserstein distance thus provides a simple and useful tool for analyzing large-ensemble datasets.
}

%TC:ignore
%TC:endignore

%TC:ignore
\maketitle
%TC:endignore

\section{Introduction}\label{sec:introduction}

Climate variability can be decomposed into forced variability and internal variability \citep{Hawkins+Sutton09,Deser+12}.
The former is driven by external forcing (e.g., radiative forcing); the latter, internal variability, results from the chaotic dynamics of the atmosphere--ocean system.
Separating these two components and quantifying their contributions are fundamental to detecting the climate change signal and assessing uncertainty in climate projections \citep{Lehner+20}.
To this end, single-model initial-condition large ensembles (SMILEs) have been developed, in which many simulations share the same climate model and external forcing but differ in their initial conditions \citep{Deser+20,Maher+21}.
In a SMILE, internal variability is assumed to average out over a sufficient number of ensemble members.
The ensemble mean is then regarded as the (externally) forced component and the residual from this mean as the internal component \citep{Frankcombe+18,Lehner+20}.
Their relative dominance can thus be assessed at each location.

Several metrics have been proposed to compare the relative importance of forced and internal variability.
These metrics are commonly based on the variance \citep{Llovel+18,Waldman+18,Yettella+18}.
One such metric expresses the relative magnitude of internal variability through the ratio of the variances (or standard deviations) of the ensemble mean and the residual \citep{Leroux+18,Nonaka+20}.
The variance fully characterizes variability only for Gaussian distributions.
Climate variables are often skewed or heavy-tailed \citep{Franzke+20}, and the variance alone may not adequately capture such non-Gaussianity.
To address this limitation, \cite{Sane+24} defined an information-theoretic metric that uses the Shannon entropy and mutual information to evaluate the magnitudes of forced and internal variability as distribution-based uncertainties.
They applied the metric to non-Gaussian variables and showed that it is more robust to outliers than the variance-based metric.
Computing their metric, however, requires binning the data into a histogram, so that the estimate depends on an additional parameter, the bin width.

In addition to information-theoretic quantities, optimal transport costs are widely used in machine learning to compare two probability distributions \citep[e.g.,][]{Arjovsky+17,Peyre+Cuturi19}.
Information theory measures how much uncertainty (or information) is shared between two variables \citep[e.g.,][]{Cover+Thomas05}.
Optimal transport, in contrast, measures a geometric difference: the minimal distance (or cost) over which one distribution must be moved to match the other \citep[e.g.,][]{Friesecke+24}.
This geometric view is useful in atmosphere--ocean science, including climatology.
Indeed, optimal transport has been used to detect changes in forced climate attractors \citep{Robin+17,Tel+20}, to evaluate the performance of climate models \citep{Vissio+20,Garrett+24,Houedry+26}, to compute ensemble means \citep{Duc+Sawada24,LeCoz+25}, to compare precipitation patterns across ensemble members \citep{Nishizawa24}, and to assimilate observational data \citep{Tamang+21,Bocquet+24}.
However, the utility of optimal transport for analyzing variability in SMILEs has not yet been fully explored.

This study proposes a metric for the relative magnitude of internal variability, based on the Wasserstein distance, a form of optimal transport cost.
The proposed metric is easily computed without additional parameters, while it incorporates the entire distribution shape, including non-Gaussianity.
``\nameref{sec:methods}'' reviews the existing metrics and then defines the proposed metric.
``\nameref{sec:data}'' describes synthetic climate data and a large-ensemble dataset, the Community Earth System Model Large Ensemble \citep[CESM-LE;][]{Kay+15}.
``\nameref{sec:results}'' demonstrates the validity of the proposed metric using the synthetic data and then applies it to the 2~m air temperature and total precipitation in CESM-LE.
The last section presents ``\nameref{sec:conclusions}.''

\section{Methods}\label{sec:methods}

\subsection{Existing internal-variability metrics}

\begin{figure*}[tb]
    \centering
    \includegraphics[width=1.0\textwidth]{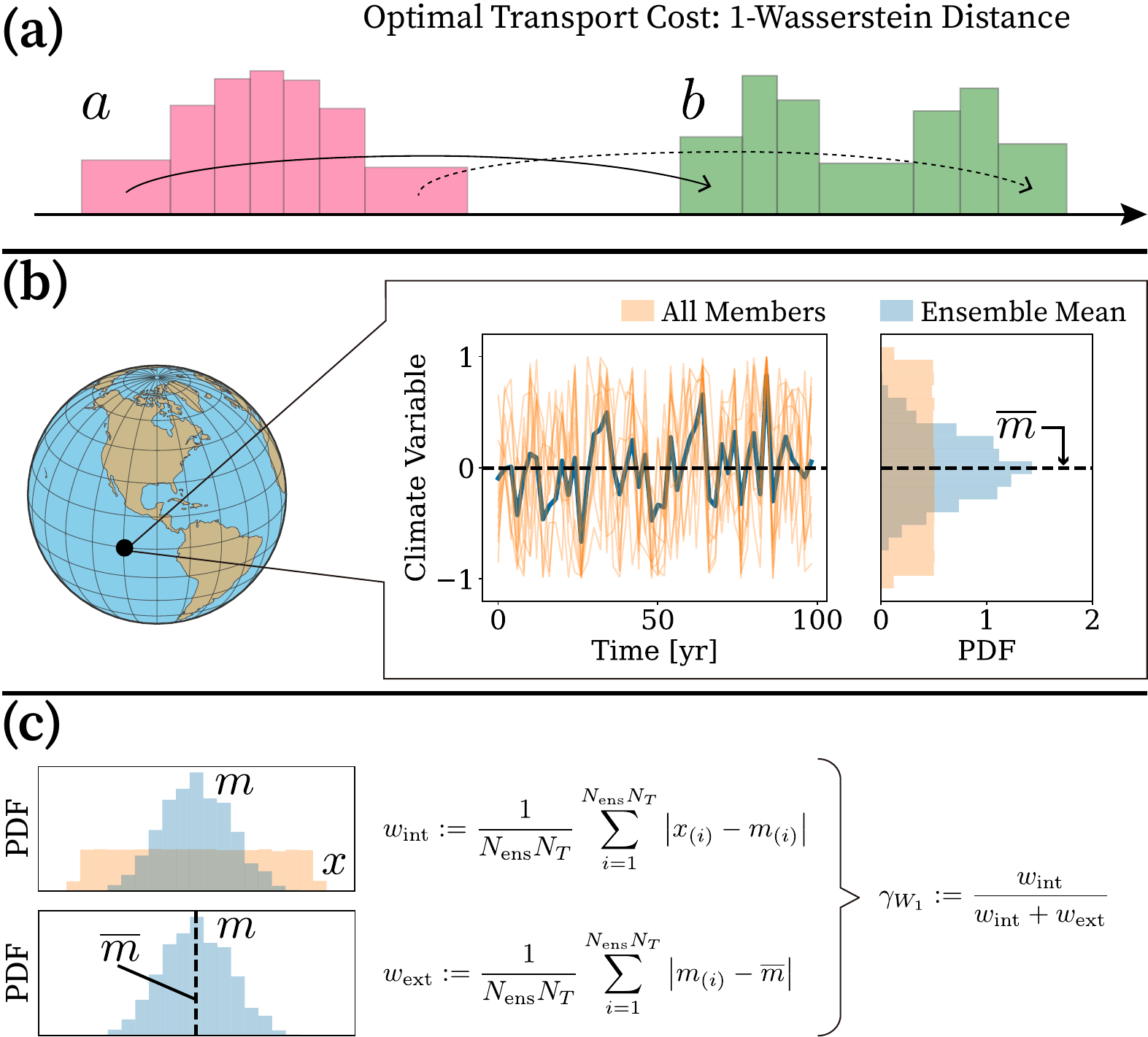}
    \caption{
        Schematic of the 1-Wasserstein distance and of the proposed metric $\gamma_{W_1}$.
        (a) The optimal transport cost between two distributions $a$ and $b$ on $\mathbb{R}^1$.
        Each distribution is drawn as a histogram with bins of equal area.
        The cost is minimized by matching the bins in order of position (arrows).
        (b) Ensemble time series of a climate variable at a single location (globe), with all members $x_{n,t}$ (orange) and the ensemble mean $m_t$ (blue).
        The samples at all time steps give the distributions on the right (probability density function; PDF).
        The dashed line indicates the total mean $\overline{m}$.
        (c) The two 1-Wasserstein distances computed from these distributions.
        Here, $w_{\mathrm{int}}$ is the distance between the distributions of $x_{n,t}$ and $m_t$ (upper row), and $w_{\mathrm{ext}}$ is that between the distribution of $m_t$ and $\overline{m}$ (lower row).
        These two distances together define $\gamma_{W_1}$ [Eqs.~(\ref{eq:sort_x})--(\ref{eq:gamma_W1})].
    }
    \label{fig01}
\end{figure*}

We first review the variance-based metric $\gamma_{\mathrm{var}}$ \citep{Waldman+18} and the information-theoretic metric $\gamma_{\mathrm{info}}$ \citep{Sane+24}.
Throughout this study, we consider ensemble time series $x_{n,t}$ of a climate variable at a single location (Fig.~\ref{fig01}b), where $n$ indexes the ensemble members ($n=1,\dots,N_{\mathrm{ens}}$) and $t$ indexes time ($t=1,\dots,N_T$).
All the metrics considered here are computed by treating the samples at all time steps as a single set.
This computation assumes that internal variability is stationary over the analysis period, whereas the forced component may vary in time (e.g., a trend).
Non-stationarity, such as the time of emergence of the forced response or a change in the statistics of internal variability \citep[e.g.,][]{Franzke+26}, could be examined by computing the metrics over rolling windows.

%TC:ignore
%TC:endignore

As preliminaries, we define the ensemble mean, the total mean, and the residual as follows:
\begin{align}
    m_t &:= \frac{1}{N_{\mathrm{ens}}} \sum_{n=1}^{N_{\mathrm{ens}}} x_{n,t},\label{eq:ensemble-mean} \\
    \overline{m} &:= \frac{1}{N_T} \sum_{t=1}^{N_T} m_t, \\
    \eta_{n,t} &:= x_{n,t} - m_t.
\end{align}
Here, the total mean $\overline{m}$ is the temporal mean of the ensemble mean $m_t$.
The variable is decomposed as $x_{n,t} = m_t + \eta_{n,t}$.
We interpret the ensemble mean $m_t$ as the forced component and the residual $\eta_{n,t}$ as the internal component \citep{Frankcombe+18,Lehner+20}, and define a ratio for their relative magnitudes.

The variance-based metric $\gamma_{\mathrm{var}}$ is defined as follows \citep{Waldman+18,Sane+24}:
\begin{align}
    \gamma_{\mathrm{var}}   &:= \frac{\sqrt{\sigma_{\mathrm{int}}^2}}{\sqrt{\sigma_{\mathrm{int}}^2+\sigma_{\mathrm{ext}}^2}},\label{eq:gamma_var} \\
    \sigma_{\mathrm{int}}^2 &:= \frac{1}{N_{\mathrm{ens}}N_T}\sum_{n,t}\eta_{n,t}^2 = \frac{1}{N_{\mathrm{ens}}N_T}\sum_{n,t}(x_{n,t}-m_t)^2,\label{eq:sigma_int} \\
    \sigma_{\mathrm{ext}}^2 &:= \frac{1}{N_T}\sum_{t} \left(m_t-\overline{m}\right)^2,\label{eq:sigma_ext}
\end{align}
where $\sigma_{\mathrm{int}}^2$ and $\sigma_{\mathrm{ext}}^2$ denote the variances of the internal and forced components, respectively.
The subscripts $\mathrm{int}$ and $\mathrm{ext}$ abbreviate ``internal'' and ``externally forced.''
Thus, $\gamma_{\mathrm{var}}$ is the ratio of the magnitude of internal variability $\sqrt{\sigma_{\mathrm{int}}^2}$ to that of the total variability $\sqrt{\sigma_{\mathrm{int}}^2+\sigma_{\mathrm{ext}}^2}$.
$\gamma_{\mathrm{var}}$ is dimensionless and lies between 0 and 1.

The information-theoretic metric $\gamma_{\mathrm{info}}$ is defined as follows \citep{Sane+24}:
\begin{align}
    \gamma_{\mathrm{info}} &:= 1-\frac{I(X;M)}{H(X)} = \frac{H(X\mid M)}{H(X)},\label{eq:gamma_info}\\
    H(X\mid M) &= H(X) - I(X;M),
\end{align}
where $I(X;M)$ denotes the mutual information and $H(X)$ the Shannon entropy \citep[e.g.,][]{Cover+Thomas05}.
Following the convention in information theory, we denote random variables by uppercase letters and their realizations by lowercase letters, so that $X$ and $M$ correspond to $x_{n,t}$ and $m_t$, respectively.
Section~S1 in the supplementary material describes how these quantities are computed.

The mutual information $I(X;M)$ measures the part of the uncertainty in $X$ that is explained by $M$.
The remaining part, the conditional Shannon entropy $H(X\mid M)$, represents the uncertainty that persists even when $M$ is known, and is interpreted as the magnitude of internal variability.
The unconditional Shannon entropy $H(X)$, in contrast, represents the total uncertainty in $X$ and is interpreted as the magnitude of the total variability.
Like $\gamma_{\mathrm{var}}$, the metric $\gamma_{\mathrm{info}}$ is the ratio of the magnitude of internal variability $H(X\mid M)$ to that of the total variability $H(X)$.
$\gamma_{\mathrm{info}}$ is dimensionless and lies between 0 and 1.

\subsection{Proposed internal-variability metric}

We now construct the proposed metric from an optimal transport cost.
This construction requires only the Wasserstein distance on the one-dimensional metric space $\mathbb{R}^1$.
For optimal transport in general, including Wasserstein distances in higher dimensions, see, e.g., \cite{Friesecke+24}.

Optimal transport expresses the difference between two distributions as the minimum cost of deforming one into the other.
Consider two probability distributions on $\mathbb{R}^1$ (Fig.~\ref{fig01}a).
If each distribution is drawn as a histogram whose bins have equal area (i.e., equal probability mass), the transport cost is the average of the distances over which the bins are moved.
The bins can be moved in many ways, but the cost is minimized by matching them in order of position, that is, by moving the $i$-th bin of one distribution to the $i$-th bin of the other \citep[e.g.,][]{Friesecke+24}.
This minimum cost is the optimal transport cost, called the Wasserstein distance when the cost is measured by a distance in the underlying metric space.
In one dimension, the histogram is used only to illustrate the transport, as the Wasserstein distance is determined by the differences between corresponding quantiles.

To obtain these quantiles, we sort all $N_{\mathrm{ens}}N_T$ values of $x_{n,t}$ in ascending order.
The values of $m_t$ are sorted in the same way, after each $m_t$ is replicated $N_{\mathrm{ens}}$ times so that the two samples have the same length:
\begin{align}
    x_{(1)} &\le x_{(2)} \le \cdots \le x_{(N_{\mathrm{ens}}N_T)},\label{eq:sort_x} \\
    m_{(1)} &\le m_{(2)} \le \cdots \le m_{(N_{\mathrm{ens}}N_T)}.\label{eq:sort_m}
\end{align}
For example, $x_{(2)}$ is the second smallest of all $x_{n,t}$.
Both $x_{(i)}$ and $m_{(i)}$ are then the $i/(N_{\mathrm{ens}}N_T)$-quantiles of their respective distributions.

Following $\gamma_{\mathrm{var}}$ [Eq.~(\ref{eq:gamma_var})], we define quantities that characterize internal and forced variability and take their ratio as the metric (Fig.~\ref{fig01}c).
The magnitude of internal variability is given by the Wasserstein distance $w_{\mathrm{int}}$ between the distribution of all samples $x_{n,t}$ and that of the ensemble mean $m_t$:
\begin{align}
    w_{\mathrm{int}} &:= \frac{1}{N_{\mathrm{ens}}N_T}\sum_{i=1}^{N_{\mathrm{ens}}N_T} \lvert x_{(i)}- m_{(i)} \rvert.\label{eq:w_int}
\end{align}
The cost here is the distance on $\mathbb{R}^1$ (i.e., the absolute difference), and thus $w_{\mathrm{int}}$ is the 1-Wasserstein distance.
The magnitude of forced variability is likewise given by the 1-Wasserstein distance between the distribution of $m_t$ and that of the total mean $\overline{m}$:
\begin{align}
    w_{\mathrm{ext}} &:= \frac{1}{N_{\mathrm{ens}}N_T}\sum_{i=1}^{N_{\mathrm{ens}}N_T} \lvert m_{(i)}-\overline{m} \rvert.\label{eq:w_ext}
\end{align}
Although $\overline{m}$ is a single value, it can be regarded as replicated $N_{\mathrm{ens}}N_T$ times, so that $w_{\mathrm{ext}}$ takes the same form as $w_{\mathrm{int}}$.

Finally, we define $\gamma_{W_1}$ from these two quantities as follows:
\begin{align}
    \gamma_{W_1} &:= \frac{w_{\mathrm{int}}}{w_{\mathrm{int}}+w_{\mathrm{ext}}}.\label{eq:gamma_W1}
\end{align}
Following $\gamma_{\mathrm{var}}$ and $\gamma_{\mathrm{info}}$, we interpret the numerator $w_{\mathrm{int}}$ as the magnitude of internal variability and the denominator $w_{\mathrm{int}}+w_{\mathrm{ext}}$ as that of the total variability.
$\gamma_{W_1}$ lies between 0 and 1, and larger values suggest a greater relative contribution of internal variability.

The denominator $w_{\mathrm{int}}+w_{\mathrm{ext}}$ in Eq.~(\ref{eq:gamma_W1}) is a sum of two transport costs and not a single one.
The distribution of all samples $x_{n,t}$ can also be transported directly to the total mean $\overline{m}$, at a single cost $w_{\mathrm{tot}} := (\sum_i \lvert x_{(i)}-\overline{m} \rvert)/(N_{\mathrm{ens}}N_T)$.
This cost satisfies $w_{\mathrm{tot}} \le w_{\mathrm{int}}+w_{\mathrm{ext}}$ by the triangle inequality, with equality only when each $m_{(i)}$ lies between $x_{(i)}$ and $\overline{m}$.
Normalizing $w_{\mathrm{int}}$ by $w_{\mathrm{tot}}$ defines a second metric, $\widetilde{\gamma}_{W_1} := w_{\mathrm{int}}/w_{\mathrm{tot}}$, which is larger than or equal to $\gamma_{W_1}$.
For the synthetic climate data and CESM-LE, the results reported below do not change when $\widetilde{\gamma}_{W_1}$ is used instead of $\gamma_{W_1}$ (Section~S2 in the supplementary material), which suggests that the difference between the two metrics is negligible in practice.

The proposed metric $\gamma_{W_1}$ combines the advantages of $\gamma_{\mathrm{var}}$ and $\gamma_{\mathrm{info}}$.
The metric $\gamma_{\mathrm{var}}$ is easy to compute because it is obtained from the differences between $x_{n,t}$ and $m_t$ [Eqs.~(\ref{eq:gamma_var})--(\ref{eq:sigma_ext})].
However, $\gamma_{\mathrm{var}}$ relies on the variance alone and thus does not adequately reflect the statistics of non-Gaussian variables.
The metric $\gamma_{\mathrm{info}}$ reflects these statistics through its information-theoretic construction, but requires histograms (Section~S1 in the supplementary material).
Constructing histograms is cumbersome and makes the metric depend on how the bins are chosen.
The proposed $\gamma_{W_1}$ is nearly as easy to compute as $\gamma_{\mathrm{var}}$, requiring only sorting and differencing and no additional parameters.
Yet it reflects the entire distribution shape, as $\gamma_{\mathrm{info}}$ does, because the underlying 1-Wasserstein distance measures the difference between any two distributions, including non-Gaussian ones.

\section{Data}\label{sec:data}

We use two datasets: synthetic climate data and CESM-LE.
The former is used to show the validity of the proposed metric $\gamma_{W_1}$.
The latter is then analyzed with all three metrics (i.e., $\gamma_{\mathrm{var}}$, $\gamma_{\mathrm{info}}$, and $\gamma_{W_1}$).
All results reported in this paper can be reproduced with our publicly available code (``\nameref{subsec:data-availability}'').

The synthetic climate data are random time series that mimic real climate data, following \cite{Sane+24}.
The random numbers were drawn from the Gaussian and uniform distributions used in their study and, in addition, from a lognormal distribution intended to represent heavy-tailed variables such as precipitation.
The time series are independent and identically distributed in time, and each experiment uses one of these three distributions.
Section~S3 in the supplementary material describes the generation procedure in detail, and Section~S4 confirms that introducing temporal correlation hardly changes the results.

To emulate forced and internal variability, we imposed a Pearson correlation coefficient $\rho$ between ensemble members and varied it over $[0,1)$.
The correlated part is regarded as the forced component and the uncorrelated part as the internal component \citep{Sane+24}.
As $\rho$ increases, the members become more similar, and forced variability increases (i.e., internal variability decreases).
The ensemble size $N_{\mathrm{ens}}$ was varied from 10 to 200.
The time step was one month, and the time series length $T$ was varied from 10 to 1,000 years (the number of time steps was $N_T = 12\,T$).

To examine robustness to outliers, we also conducted experiments in which a fraction of the samples was replaced by artificial outliers.
Following \cite{Sane+24}, 0.5\% of all samples were replaced by a value of 5.0 for the Gaussian distribution and by 1.5 for the uniform distribution.
The former value is five standard deviations from the mean, and the latter lies outside the support $[-1,1]$ (Section~S3 in the supplementary material).
For the lognormal distribution, the outlier value was 16.6, which corresponds to the 99.75th percentile.
Every experiment was repeated with five random seeds, and the metrics were averaged over these five realizations before any further analysis.

As an application to a large ensemble, we used the 2~m air temperature and total precipitation from CESM-LE \citep{Kay+15,Beaujardiere+19}.
CESM-LE is a 40-member initial-condition ensemble of coupled atmosphere--ocean simulations under identical radiative forcing, with a grid spacing of 1.25$^{\circ}$ in longitude and about 0.94$^{\circ}$ in latitude.
We analyzed the monthly mean output and removed the seasonal cycle by subtracting the monthly climatology common to all ensemble members \citep{Wills+20}.
The trend associated with external forcing is part of the forced component, so we did not apply linear detrending \citep{Sane+24}.
We analyzed the model outputs from two forcing experiments: one under 20th-century historical forcing (20C) for 1920--2005 and the other under representative concentration pathway 8.5 \citep[RCP8.5;][]{Meinshausen+11} for 2006--2100.
For each set, the three metrics ($\gamma_{\mathrm{var}}$, $\gamma_{\mathrm{info}}$, and $\gamma_{W_1}$) were computed independently at each grid point.
Comparing the resulting maps, we show how the relative magnitude of internal variability differs between the two forcing experiments.
\section{Results}\label{sec:results}

Figure~\ref{fig01} is a schematic, but the time series shown there are synthetic climate data based on the uniform distribution with $\rho=0.25$, $N_{\mathrm{ens}}=200$, and $T=100$~years.
The samples of all ensemble members $x_{n,t}$ vary around the ensemble mean $m_t$, reflecting the correlation between members.
Using such synthetic data, we examine the sensitivity of each metric to outliers, to the ensemble size $N_{\mathrm{ens}}$, and to the time series length $T$.

\begin{figure*}[tb]
    \centering
    \includegraphics[width=1.0\textwidth]{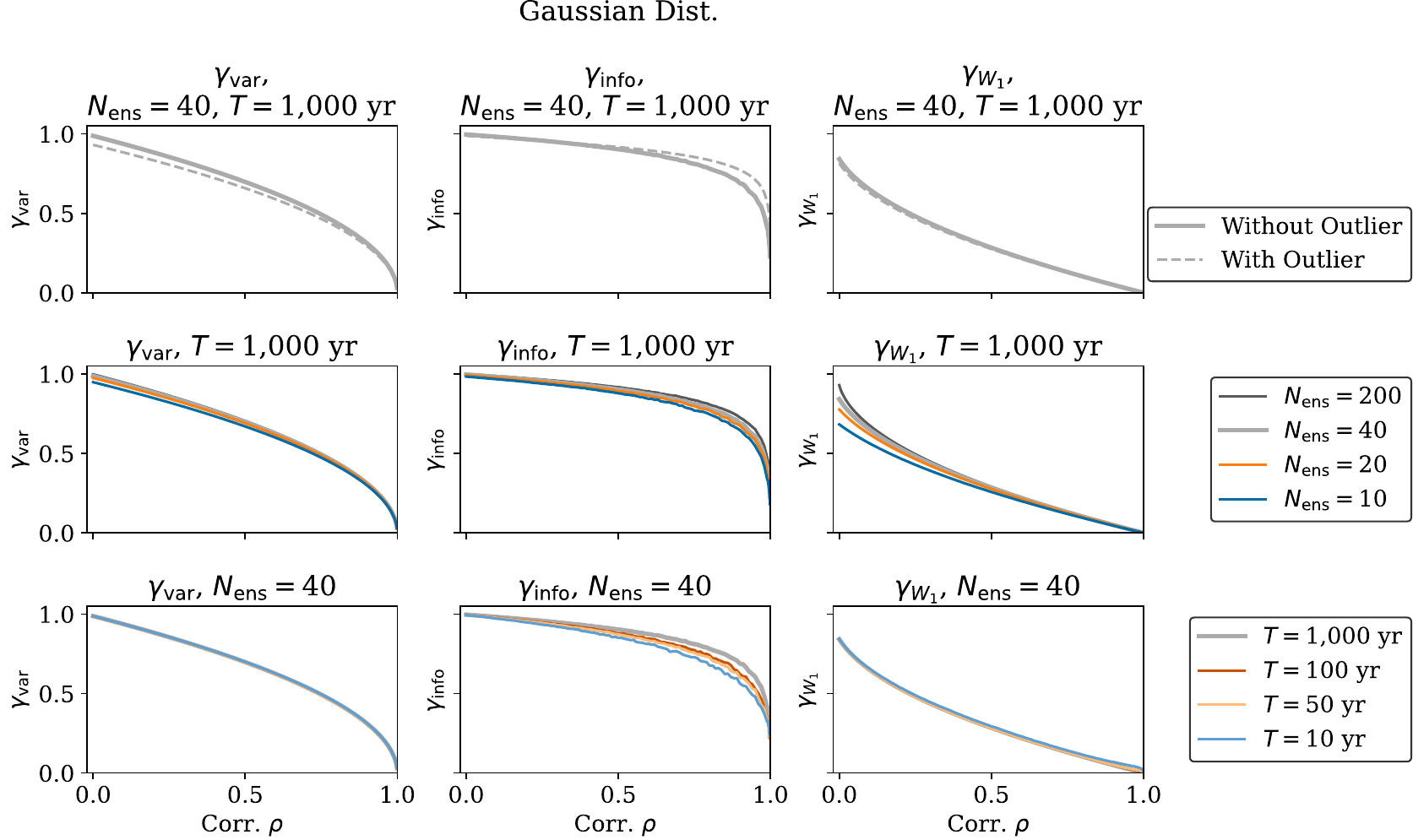}
    \caption{
        Sensitivity of the three metrics for synthetic data based on the Gaussian distribution.
        Each column shows one metric ($\gamma_{\mathrm{var}}$, $\gamma_{\mathrm{info}}$, and $\gamma_{W_1}$, from left to right) as a function of the correlation coefficient $\rho$.
        Top: with (dashed) and without (solid) outliers, for $N_{\mathrm{ens}}=40$ and $T=1{,}000$~years.
        Middle and bottom: four values of the ensemble size $N_{\mathrm{ens}}$ ($T=1{,}000$~years) and of the data (time series) length $T$ ($N_{\mathrm{ens}}=40$), respectively, both without outliers.
    }
    \label{fig02}
\end{figure*}

Figure~\ref{fig02} shows the results for the synthetic data based on the Gaussian distribution.
From top to bottom, the figure presents the dependence on outliers, $N_{\mathrm{ens}}$, and $T$.
All three metrics ($\gamma_{\mathrm{var}}$, $\gamma_{\mathrm{info}}$, and $\gamma_{W_1}$) decrease as the correlation coefficient $\rho$ increases.
This decrease is consistent with the design of the synthetic data (``\nameref{sec:data}''), in which a larger $\rho$ narrows the spread among members and thus reduces the relative contribution of internal variability.
The metrics, however, differ in where they decrease steeply: $\gamma_{W_1}$ near $\rho = 0$, and $\gamma_{\mathrm{var}}$ and $\gamma_{\mathrm{info}}$ near $\rho = 1$, as we discuss later.
Section~S5 in the supplementary material gives the corresponding results for the uniform and lognormal distributions (Figs.~\ref{figS04} and~\ref{figS05}).

%TC:ignore
%TC:endignore

For a quantitative evaluation, we defined the reference as the values obtained without outliers and with the largest sample size ($N_{\mathrm{ens}} = 200$ and $T = 1{,}000$ years).
We then measured the departure from this reference by integrating the absolute difference over $\rho \in [0,1)$.
To examine the convergence of the reference, we further increased $N_{\mathrm{ens}}$ or $T$ (Section~S6 in the supplementary material).
Both $\gamma_{\mathrm{var}}$ and $\gamma_{W_1}$ converged sufficiently: their departure from the reference remained small.
The metric $\gamma_{\mathrm{info}}$ did not converge, because the bin width estimated from the data changes with the sample size.
This explanation is supported by a theoretical argument showing that $\gamma_{\mathrm{info}}$ depends on the bin width through the Shannon entropy (Section~S7 in the supplementary material).

%TC:ignore
\begin{figure*}[tb]
    \centering
    \includegraphics[width=1.0\textwidth]{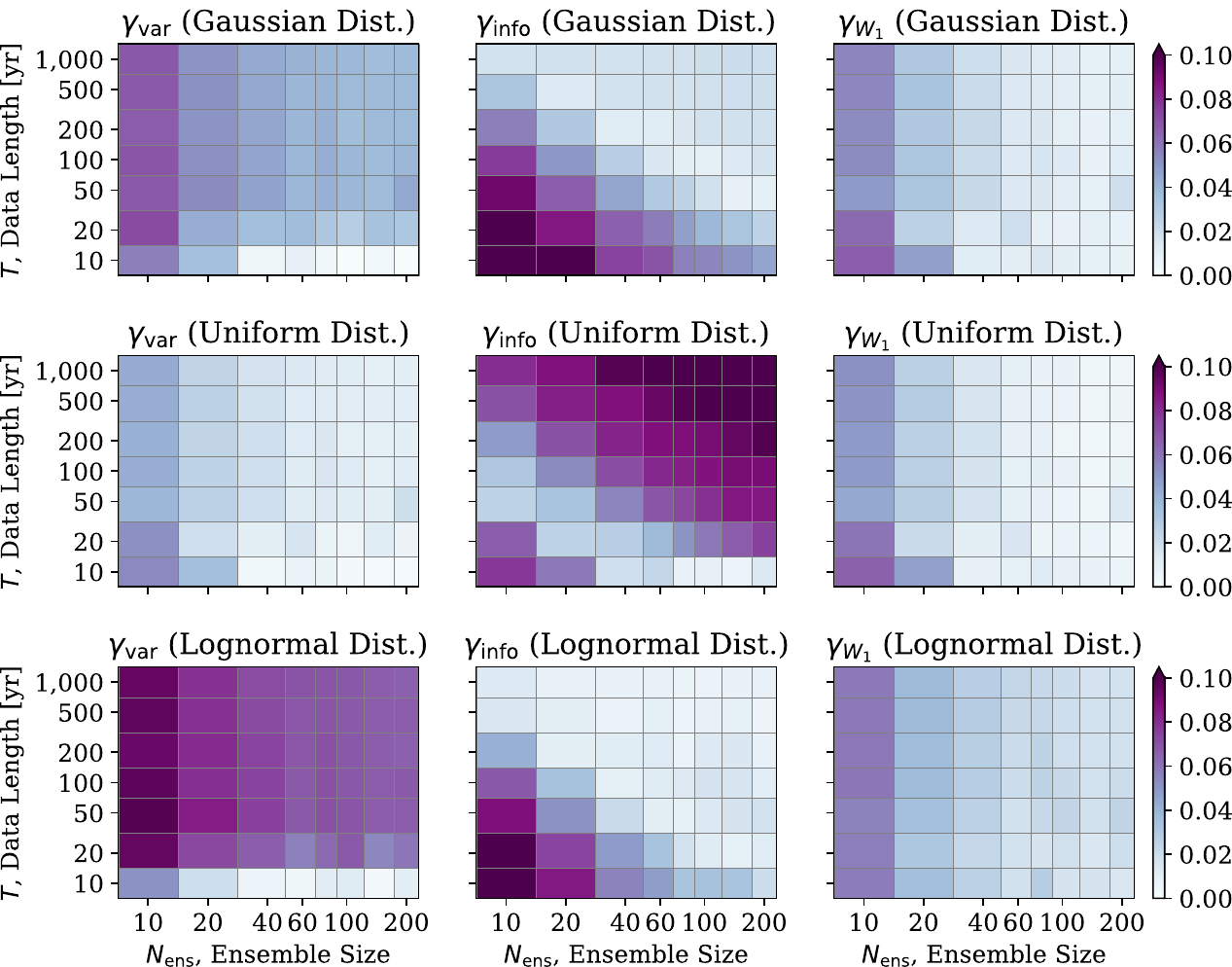}
    \caption{
        Departure of the three metrics from the reference for synthetic data with outliers.
        Columns show the metrics ($\gamma_{\mathrm{var}}$, $\gamma_{\mathrm{info}}$, and $\gamma_{W_1}$, from left to right) and rows the Gaussian, uniform, and lognormal distributions.
        Each panel gives the departure on a grid of the ensemble size $N_{\mathrm{ens}}$ and the data (time series) length $T$.
        The reference is the value obtained without outliers for $N_{\mathrm{ens}}=200$ and $T=1{,}000$~years, and the departure is the absolute difference from it integrated over $\rho \in [0,1)$.
    }
    \label{fig03}
\end{figure*}
%TC:endignore

Figure~\ref{fig03} shows the departure from the reference on a grid of $N_{\mathrm{ens}}$ and $T$ in the presence of outliers (for the case without outliers, see Section~S5 in the supplementary material).
The metric $\gamma_{\mathrm{var}}$ is sensitive to outliers but nearly independent of $T$ (Figs.~\ref{fig03} and~\ref{figS06}), in agreement with \cite{Sane+24}.
Contrary to their finding, $\gamma_{\mathrm{info}}$ is also sensitive to outliers.
This difference is due to our treatment of the histogram binning required to compute $\gamma_{\mathrm{info}}$, namely the number of bins and the range of values, which we estimated from the data.
The dependence on the binning also explains the sensitivity of $\gamma_{\mathrm{info}}$ to $N_{\mathrm{ens}}$ and $T$.
In their source code, the binning is fixed \citep{Sane+25}.
As they discuss, however, it should in principle be determined from the data \citep{Sane+24}, so we followed the procedure in their paper.
When we fixed it as well, all the dependences weakened markedly and we reproduced their results (not shown).
For some shapes of the distribution, the departure of $\gamma_{\mathrm{info}}$ and $\gamma_{\mathrm{var}}$ remains large even for large $N_{\mathrm{ens}}$ and $T$ (Fig.~\ref{fig03}), which implies that their estimates are unstable.

In contrast, the proposed metric $\gamma_{W_1}$ is largely insensitive to outliers and to $T$, although it does depend on $N_{\mathrm{ens}}$ (Figs.~\ref{fig03} and~\ref{figS06}).
The departure of $\gamma_{W_1}$ is small overall and largely independent of the distribution shape.
It becomes large only for $N_{\mathrm{ens}} \lesssim 20$ (up to 0.07 for $N_{\mathrm{ens}} = 10$) and reflects the sampling error of the ensemble mean $m_t$, which scales as $1/\sqrt{N_{\mathrm{ens}}}$.
For $N_{\mathrm{ens}} \ge 40$, it is at most 0.03 for all $T$ and distributions, with or without outliers.
Therefore, for $N_{\mathrm{ens}} \sim 40$ or more, $\gamma_{W_1}$ gives stable estimates in the presence of outliers irrespective of the distribution shape.
This threshold of $N_{\mathrm{ens}}$ may provide a guideline for choosing the ensemble size when designing large-ensemble simulations.

We also varied the fraction of outliers from 0.1\% to 5\% and measured the departure from the case without outliers (Section~S8 in the supplementary material).
The departure of $\gamma_{W_1}$ increases gradually with the fraction and remains smaller than that of $\gamma_{\mathrm{var}}$.
This result hardly depends on $N_{\mathrm{ens}}$ and $T$.
The departure of $\gamma_{\mathrm{info}}$, in contrast, can be large even for the smallest fraction, because the outliers can change the bin width estimated from the data.

These results indicate that $\gamma_{W_1}$ is well suited to analyzing the relative magnitude of internal variability in large-ensemble datasets.
Table~\ref{table01} summarizes the characteristics of the three metrics.
The main requirement for using $\gamma_{W_1}$ is a sufficient ensemble size: the metric is reliable once $N_{\mathrm{ens}}$ reaches about 40.
We revisit this dependence on $N_{\mathrm{ens}}$ in the analysis of CESM-LE.
In terms of computational complexity, $\gamma_{\mathrm{info}}$ is the most expensive, followed by $\gamma_{W_1}$ and then $\gamma_{\mathrm{var}}$.
Computing $\gamma_{\mathrm{info}}$ requires a search over candidate bin numbers, which gives $O(NB)$, where $N = N_{\mathrm{ens}}N_T$ is the total sample size and $B = 500$ is the number of candidates searched (Section~S1 in the supplementary material).
Neither $\gamma_{\mathrm{var}}$ nor $\gamma_{W_1}$ involves such additional parameters, so no search is needed.
The metric $\gamma_{\mathrm{var}}$ is based on the variance, giving $O(N)$, whereas $\gamma_{W_1}$ requires sorting the samples, giving $O(N \log N)$ with quicksort.

%TC:ignore
\begin{table}[tb]
    \centering
    \caption{
        Characteristics of the three internal-variability metrics.
        ``Binning'' means the number of bins and their range, required only by $\gamma_{\mathrm{info}}$.
        $N := N_{\mathrm{ens}}N_T$, and $B = 500$ is the largest number of bins searched.
    }
    \label{table01}
    \begin{tabular}{@{}llll@{}}
    \toprule
    Metric & Mathematical background & Additional parameters & Computational complexity \\
    \midrule
    $\gamma_{\mathrm{var}}$ & Descriptive statistics & None & $O(N)$ \\
    $\gamma_{\mathrm{info}}$ & Information theory & Binning & $O(NB)$ \\
    $\gamma_{W_1}$ & Optimal transport & None & $O(N \log N)$ \\
    \botrule
    \end{tabular}
\end{table}
%TC:endignore

The departure in Fig.~\ref{fig03} integrates the difference over $\rho$ and thus does not show the dependence of each metric on $\rho$.
The relative contribution of internal variability, which $\rho$ controls in the synthetic data, differs widely among climate variables and locations on the globe \citep{Deser+12,Lehner+20}.
Interpreting such differences with a metric requires knowledge of its dependence on $\rho$.
For all three distributions (Figs.~\ref{fig02}, \ref{figS04}, \ref{figS05}, and~\ref{figS09}), $\gamma_{W_1}$ changes rapidly with $\rho$ near $\rho = 0$, where internal variability dominates.
The other two metrics, $\gamma_{\mathrm{var}}$ and $\gamma_{\mathrm{info}}$, change rapidly near $\rho = 1$, where forced variability dominates.
Thus, $\gamma_{W_1}$ and the other two metrics have their steepest slopes at opposite ends of the range of $\rho$ (Section~S9 in the supplementary material).
For the Gaussian distribution, we also show this contrast between $\gamma_{W_1}$ and $\gamma_{\mathrm{var}}$ theoretically (Section~S10 in the supplementary material).

Figure~\ref{fig04} shows the results for the 2~m air temperature from CESM-LE.
Under the historical forcing (20C), all metrics are high over most of the globe, indicating that internal variability is dominant.
Under the RCP8.5 forcing, all metrics are lower than under 20C, indicating a smaller relative contribution of internal variability during the RCP8.5 period.
To compare the two forcing experiments, we took the difference in each metric between RCP8.5 and 20C.
The decrease is larger over the tropical to subtropical oceans, where internal variability is small, than over the Northern Hemisphere continents, where internal variability is large.
This contrast is consistent with the known geographical distribution of internal variability in temperature \citep{Deser+12,Lehner+20}, suggesting that the proposed metric captures this pattern.

%TC:ignore
\begin{figure*}[tb]
    \centering
    \includegraphics[width=1.0\textwidth]{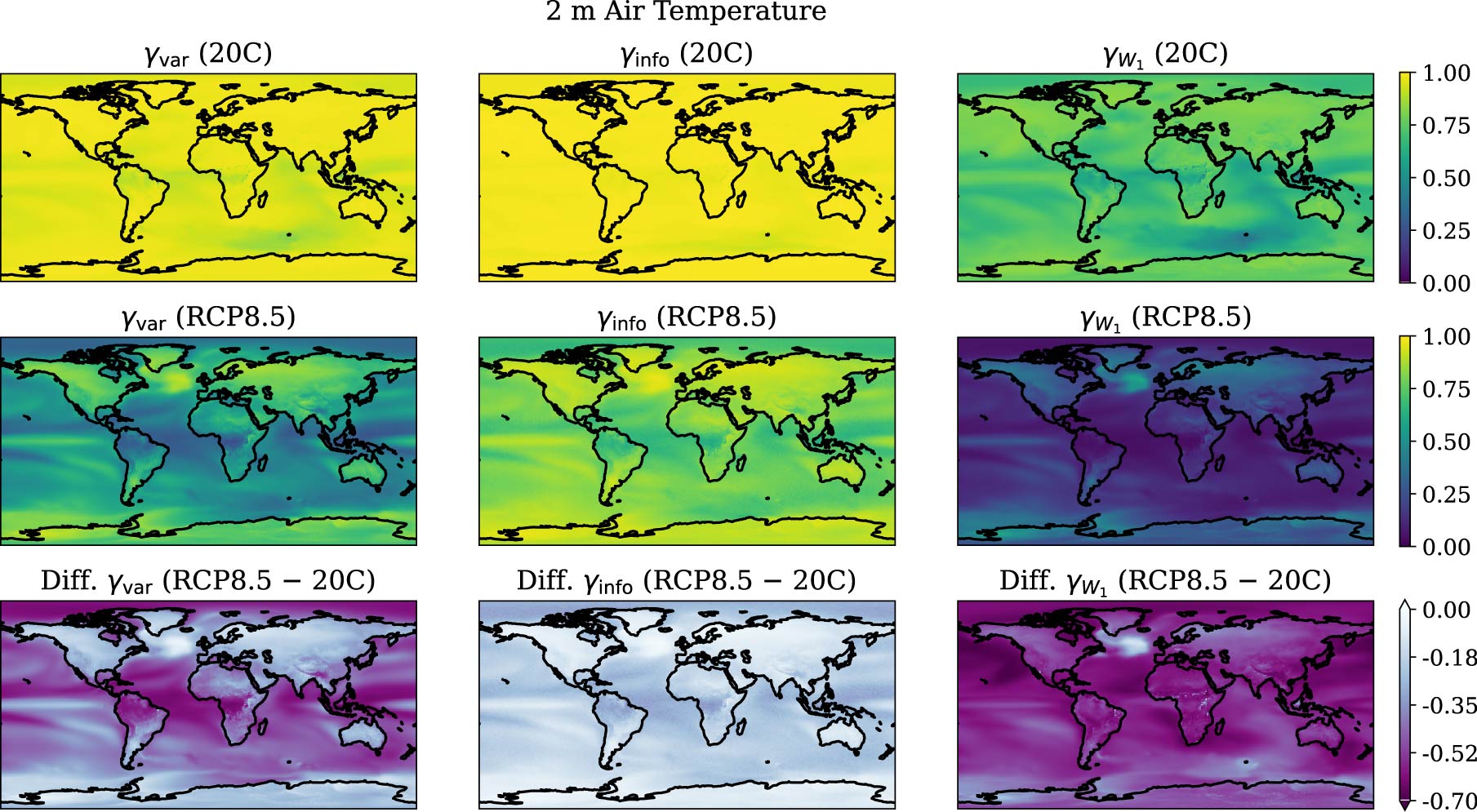}
    \caption{
        Internal-variability metrics for the 2~m air temperature from CESM-LE.
        Columns show the metrics ($\gamma_{\mathrm{var}}$, $\gamma_{\mathrm{info}}$, and $\gamma_{W_1}$, from left to right).
        Top and middle rows: the metric under 20th-century historical forcing (20C) and under RCP8.5.
        Bottom row: the difference in each metric (RCP8.5 minus 20C).
    }
    \label{fig04}
\end{figure*}
%TC:endignore

All three metrics give a similar spatial pattern of the difference but differ markedly in amplitude (the bottom row in Fig.~\ref{fig04}).
The proposed metric $\gamma_{W_1}$ shows the largest decrease, and $\gamma_{\mathrm{info}}$ the smallest.
These differences in amplitude are consistent with the dependence of each metric on $\rho$ (Figs.~\ref{fig02} and~\ref{figS09}).
Under the 20C forcing, $\gamma_{\mathrm{var}}$ is close to 1 over most of the globe.
This situation is analogous to the low-$\rho$ regime of the synthetic experiments, where internal variability dominates and $\gamma_{W_1}$ changes more rapidly than $\gamma_{\mathrm{var}}$.
Under the RCP8.5 forcing, $\gamma_{\mathrm{var}}$ decreases markedly over the tropical to subtropical oceans, which can be interpreted as a shift toward larger $\rho$.
At larger $\rho$, $\gamma_{W_1}$ changes more slowly than $\gamma_{\mathrm{var}}$.
The metric $\gamma_{\mathrm{info}}$, which shows the smallest decrease, varies little for $\rho \lesssim 0.5$ in the synthetic data (Fig.~\ref{fig02}), as also discussed by \cite{Sane+24}.

For $\gamma_{W_1}$, we further confirmed that the spatial patterns are nearly unchanged when only the first or the last 20 of the 40 members are used (details not shown).
Note that the uncertainty and the statistical significance of the difference in each metric could also be assessed by a bootstrap that resamples the ensemble members with replacement \citep[e.g.,][]{Mudelsee14}.

\begin{figure*}[tb]
    \centering
    \includegraphics[width=1.0\textwidth]{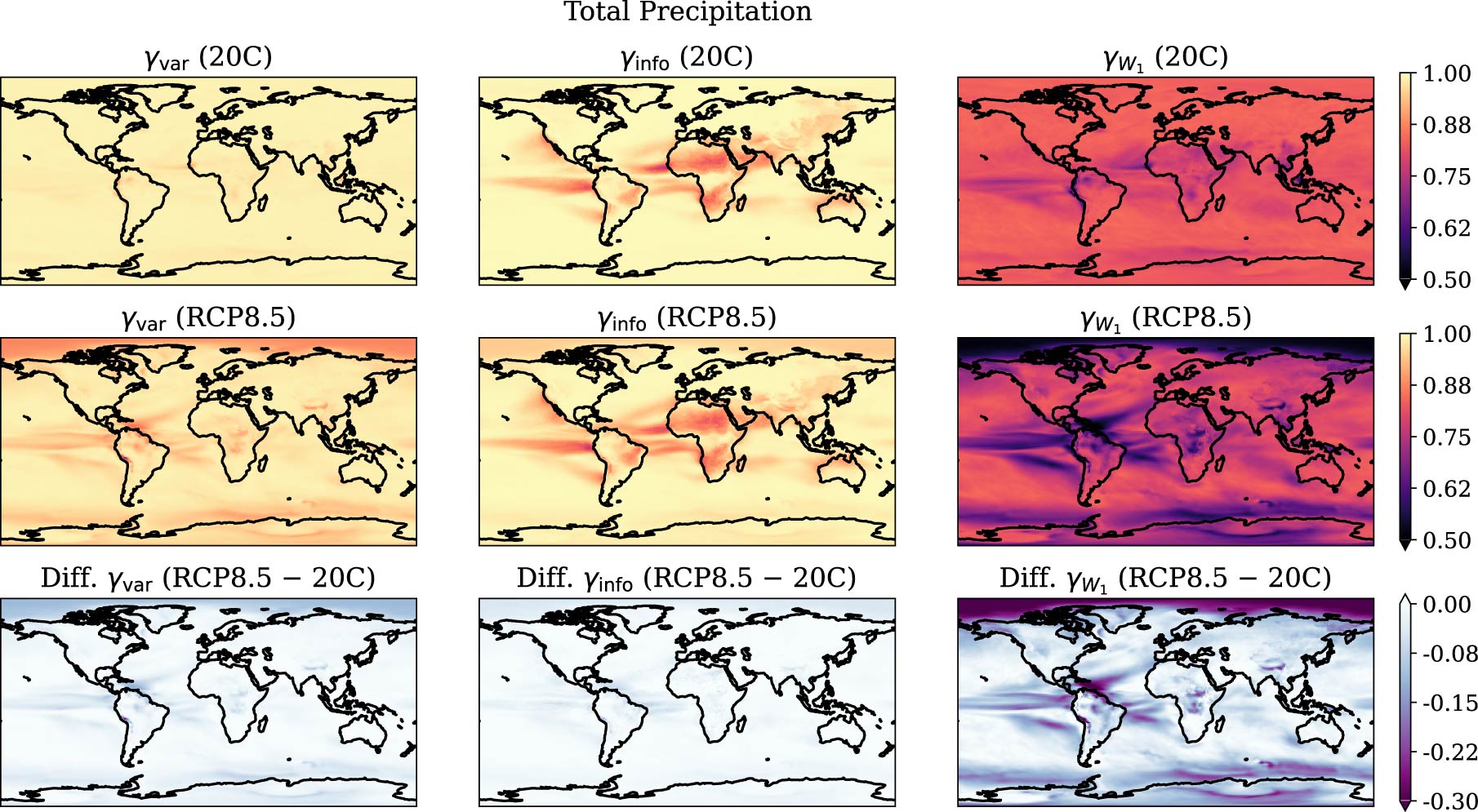}
    \caption{
        As in Fig.~\ref{fig04}, but for the total precipitation.
        Note that the color scales differ from those in Fig.~\ref{fig04}.
    }
    \label{fig05}
\end{figure*}

Figure~\ref{fig05} shows the results for the total precipitation from CESM-LE.
For precipitation, the relative contribution of internal variability tends to be large \citep{Deser+12,Lehner+20}.
Consistent with this tendency, $\gamma_{\mathrm{var}}$ is close to 1 over most of the globe under the 20C forcing.
This resembles the low-$\rho$ regime of the synthetic experiments, where internal variability dominates.
The metric $\gamma_{\mathrm{info}}$ shows spatial structure, for example along the Intertropical Convergence Zone (ITCZ) and over the continents.
The metric $\gamma_{W_1}$ takes values below 1 over many regions.
Under the RCP8.5 forcing, $\gamma_{\mathrm{var}}$ becomes lower over regions such as the Arctic, the Southern Ocean, and the tropical Pacific and Atlantic.
The difference in $\gamma_{W_1}$ (RCP8.5 minus 20C) has clear spatial structure, indicating a reduced relative contribution of internal variability at high latitudes (e.g., the Arctic) and in the tropical Atlantic and eastern Pacific.
The differences in $\gamma_{\mathrm{var}}$ and $\gamma_{\mathrm{info}}$ show spatial patterns similar to that in $\gamma_{W_1}$ but with smaller amplitudes.
As for the 2~m air temperature, the larger difference in $\gamma_{W_1}$ is consistent with $\gamma_{\mathrm{var}}$ being close to 1, as in the low-$\rho$ regime, where $\gamma_{W_1}$ changes more rapidly than the other two metrics.

%TC:ignore
%TC:endignore
\section{Conclusions}\label{sec:conclusions}

We proposed the metric $\gamma_{W_1}$, based on the 1-Wasserstein distance, to quantify the relative magnitude of internal variability at each location in a SMILE.
The metric requires only sorting and differencing, involves no additional parameters, and is applicable to non-Gaussian variables.
Using the synthetic climate data, we showed that, in the presence of outliers, $\gamma_{W_1}$ gives stable estimates irrespective of the distribution shape, unlike the existing metrics $\gamma_{\mathrm{var}}$ and $\gamma_{\mathrm{info}}$ \citep{Waldman+18,Sane+24}.
For reliable estimation, $\gamma_{W_1}$ requires an ensemble size of about 40 or more.
We also showed that $\gamma_{W_1}$ changes most rapidly when internal variability dominates, whereas $\gamma_{\mathrm{var}}$ and $\gamma_{\mathrm{info}}$ change most rapidly when forced variability dominates.
We then applied the three metrics to CESM-LE, where all of them indicate a smaller relative contribution of internal variability under RCP8.5 than under 20C, with similar spatial patterns but different amplitudes.
These differences in amplitude are consistent with the differences in the rate of change found in the synthetic data.
A detailed analysis of CESM-LE with $\gamma_{W_1}$, including the causes of internal and forced variability, is left for future work.

There are at least three directions for extending our study.
First, optimal transport also defines barycenters, which may provide an alternative to the arithmetic mean (or sample mean) for the ensemble mean \citep{Duc+Sawada24}.
Second, in addition to internal and forced variability, the uncertainty in climate projections includes model uncertainty.
Assessing it therefore requires multi-model rather than single-model ensembles \citep{Deser+20,Lehner+20}.
Third, a Wasserstein stability analysis of probability distributions may make it possible to detect the climate change signal with only a few members \citep{Xie+25}.
Each of these directions could be combined with the proposed metric $\gamma_{W_1}$.
We expect the Wasserstein distance to develop further as a practical tool for analyzing large-ensemble datasets.

%TC:ignore
\backmatter

\bmhead{Abbreviations}
\begin{description}
    \item[\textbf{CESM-LE}] Community Earth System Model Large Ensemble
    \item[\textbf{ITCZ}] Intertropical Convergence Zone 
    \item[\textbf{SMILE}] Single-Model Initial-condition Large Ensemble
    \item[\textbf{RCP8.5}] Representative Concentration Pathway 8.5
    \item[\textbf{20C}] 20th-Century historical forcing
\end{description}

\bmhead{Author contributions}
Conceptualization, Y.Y. (lead) and S.K. (supporting); methodology, Y.Y. (lead) and S.K. (supporting); software, Y.Y. (lead) and S.K. (supporting); formal analysis, Y.Y. (lead) and S.K. (supporting); data curation, Y.Y.; visualization, Y.Y.; writing -- original draft, Y.Y.; writing -- review and editing, Y.Y. and S.K.; supervision and project administration, Y.Y.; funding acquisition, Y.Y. and S.K.
Both authors read and approved the final manuscript.

\bmhead{Funding}
Y. Yasuda was supported by Japan Society for the Promotion of Science (JSPS) KAKENHI (JP26K07197). S. Kido was supported by JSPS KAKENHI (JP21K13997, JP23H01250, JP25K01073).

\bmhead{Data availability}\label{subsec:data-availability}
The CESM-LE data are available on Amazon Web Services (\url{https://doi.org/10.26024/wt24-5j82}) \citep{Beaujardiere+19}.
The source code that supports the findings of this study is preserved in the Zenodo repository (\url{https://doi.org/10.5281/zenodo.24244686}) and developed openly in the GitHub repository (\url{https://github.com/YukiYasuda2718/wasserstein_internal_variability_metric}).
These repositories include the scripts for downloading the CESM-LE data, generating the synthetic climate data, and analyzing both datasets.

\section*{Declarations}

\bmhead{Competing interests}
The authors declare no competing interests.
%TC:endignore

\begin{appendices}
\section*{Supplementary material}\label{appendix}
\setcounter{figure}{0}
\renewcommand{\thefigure}{S\arabic{figure}}
% hyperref names figure anchors by \theHfigure; without the S they collide
% with those of Figs. 1-5 in the main text.
\renewcommand{\theHfigure}{S\arabic{figure}}

\subsection*{S1 Computation of the information-theoretic metric $\gamma_{\mathrm{info}}$}

The computation of $\gamma_{\mathrm{info}}$ follows \cite{Sane+24}.
$\gamma_{\mathrm{info}}$ is obtained from the mutual information $I(X;M)$ and the Shannon entropy $H(X)$ [Eq.~(\ref{eq:gamma_info})], both of which are computed from the joint histogram of $x_{n,t}$ and $m_t$.

To construct the histogram, we first determine its range and the number of bins.
The range is taken from the minimum to the maximum of $x_{n,t}$.
Since $m_t$ is the ensemble mean of $x_{n,t}$ [Eq.~(\ref{eq:ensemble-mean})], all values of $m_t$ fall within this range.
The bins are of equal width, and their number is determined by maximum a posteriori estimation \citep{Knuth19}.
Setting the upper limit of the search to $B = 500$ bins, we compute the posterior probability at each candidate number of bins from all samples $x_{n,t}$, and then adopt the number that maximizes the posterior \citep{Sane+24}.
The histogram of $x_{n,t}$ is evaluated for each candidate, and thus the total computational complexity is $O(NB)$, where $N = N_{\mathrm{ens}} N_T$ (see Table~\ref{table01}).

We then construct the histogram, which approximates the joint probability distribution $p(X,M)$ of $x_{n,t}$ and $m_t$.
The range and the number of bins determined above are used for both axes (i.e., for $x_{n,t}$ and for $m_t$).
Denoting the bins for $x_{n,t}$ and $m_{t}$ by the indices $i$ and $j$, respectively, we approximate $p(X,M)$ by $p_{ij}$.
The marginal distributions $p(X)$ and $p(M)$ are then given by $p_i := \sum_j p_{ij}$ and $p_j := \sum_i p_{ij}$, respectively.

The mutual information $I(X;M)$ and the Shannon entropy $H(X)$ are computed as follows:
\begin{align*}
    I(X;M) &= \sum_{ij} \left[ p_{ij} \log_2 \left(\frac{p_{ij}}{p_i p_j} \right) \right], \\
    H(X) &= - \sum_{i} \left( p_i \log_2 p_i \right).
\end{align*}
The metric $\gamma_{\mathrm{info}}$ follows from these quantities \citep{Sane+24}:
\begin{align*}
    \gamma_{\mathrm{info}} &= 1-\frac{I(X;M)}{H(X)}.
\end{align*}

\subsection*{S2 Additivity of the two transport costs $w_{\mathrm{int}}$ and $w_{\mathrm{ext}}$}

We restate the two transport costs and the proposed metric [Eqs.~(\ref{eq:w_int})--(\ref{eq:gamma_W1})]:
\begin{align*}
    w_{\mathrm{int}} &= \frac{1}{N_{\mathrm{ens}}N_T}\sum_{i=1}^{N_{\mathrm{ens}}N_T} \lvert x_{(i)}- m_{(i)} \rvert, \\
    w_{\mathrm{ext}} &= \frac{1}{N_{\mathrm{ens}}N_T}\sum_{i=1}^{N_{\mathrm{ens}}N_T} \lvert m_{(i)}-\overline{m} \rvert, \\
    \gamma_{W_1} &= \frac{w_{\mathrm{int}}}{w_{\mathrm{int}}+w_{\mathrm{ext}}}.
\end{align*}
Transporting the distribution of all samples $x_{n,t}$ directly to the total mean $\overline{m}$ gives a third transport cost, from which another metric follows:
\begin{align*}
    w_{\mathrm{tot}} &= \frac{1}{N_{\mathrm{ens}}N_T}\sum_{i=1}^{N_{\mathrm{ens}}N_T} \lvert x_{(i)}-\overline{m} \rvert, \\
    \widetilde{\gamma}_{W_1} &= \frac{w_{\mathrm{int}}}{w_{\mathrm{tot}}}.
\end{align*}

The triangle inequality gives $w_{\mathrm{tot}} \le w_{\mathrm{int}}+w_{\mathrm{ext}}$.
This inequality means that the cost of transporting $x_{n,t}$ directly to $\overline{m}$ is smaller than or equal to the total cost of the two-step transport, where $x_{n,t}$ is transported to $m_t$ and then $m_t$ to $\overline{m}$.
The two metrics are therefore ordered as follows:
\begin{align*}
    \gamma_{W_1} = \frac{w_{\mathrm{int}}}{w_{\mathrm{int}}+w_{\mathrm{ext}}} \le \frac{w_{\mathrm{int}}}{w_{\mathrm{tot}}} = \widetilde{\gamma}_{W_1}.
\end{align*}
Equality holds in both relations if and only if $m_{(i)}$ lies between $x_{(i)}$ and $\overline{m}$ for every $i$, that is, if $x_{(i)}-m_{(i)}$ and $m_{(i)}-\overline{m}$ share a sign.
This condition is satisfied when the quantiles of $m_t$ are those of $x_{n,t}$ contracted toward $\overline{m}$, as in the synthetic climate data based on the uniform distribution.

In this study, we adopt $\gamma_{W_1}$ rather than $\widetilde{\gamma}_{W_1}$ for two reasons.
First, the denominator of $\gamma_{W_1}$ separates into an internal and a forced part, as those of $\gamma_{\mathrm{var}}$ and $\gamma_{\mathrm{info}}$ do, whereas $w_{\mathrm{tot}}$ does not.
This construction also bounds $\gamma_{W_1}$ between 0 and 1.
Second, $\gamma_{W_1}$ never exceeds $\widetilde{\gamma}_{W_1}$, and it is thus the conservative choice for the relative contribution of internal variability.

\begin{figure*}[tb]
    \centering
    \includegraphics[width=1.0\textwidth]{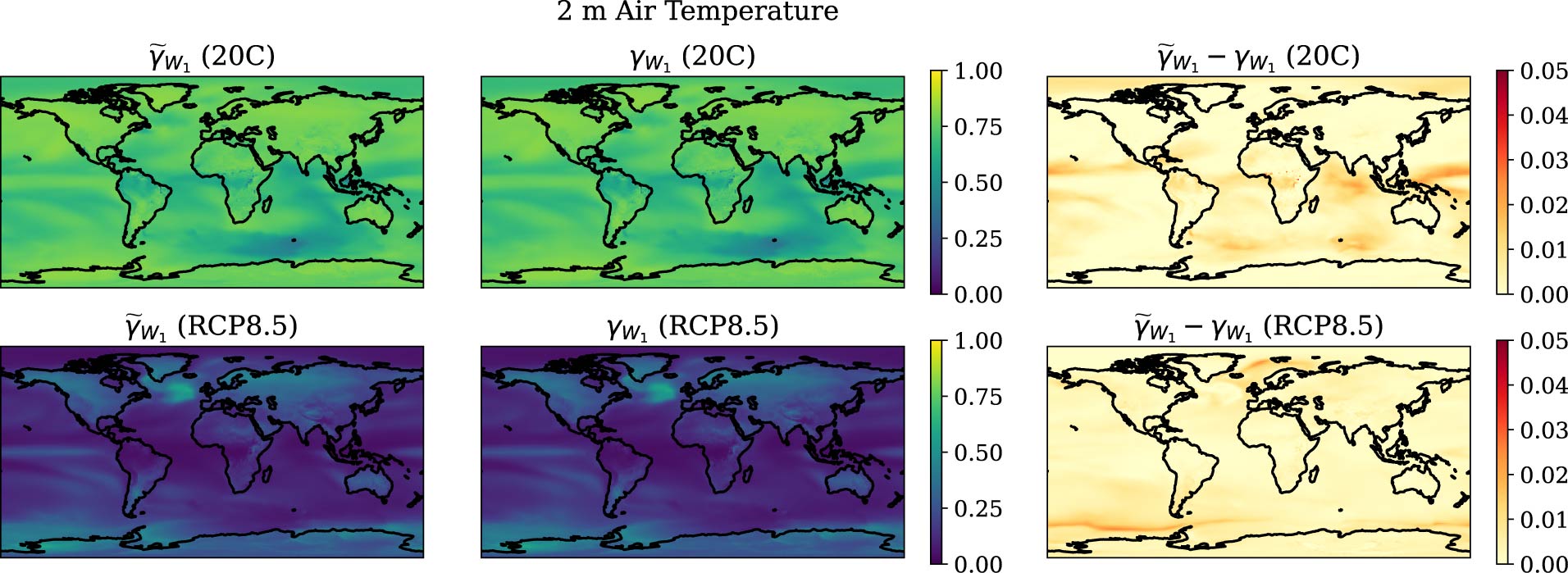}
    \caption{
        Two normalizations of the proposed metric for the 2~m air temperature from CESM-LE.
        Columns show $\widetilde{\gamma}_{W_1}$, which is normalized by the direct transport cost $w_{\mathrm{tot}}$, then $\gamma_{W_1}$, and then the difference between the two ($\widetilde{\gamma}_{W_1}-\gamma_{W_1}$).
        The middle column reproduces the $\gamma_{W_1}$ panels of Fig.~\ref{fig04} for comparison.
        Top and bottom rows: 20th-century historical forcing (20C) and representative concentration pathway 8.5 (RCP8.5).
        Both metrics are computed independently at each grid point from the same 40 ensemble members.
    }
    \label{figS01}
\end{figure*}

\begin{figure*}[tb]
    \centering
    \includegraphics[width=1.0\textwidth]{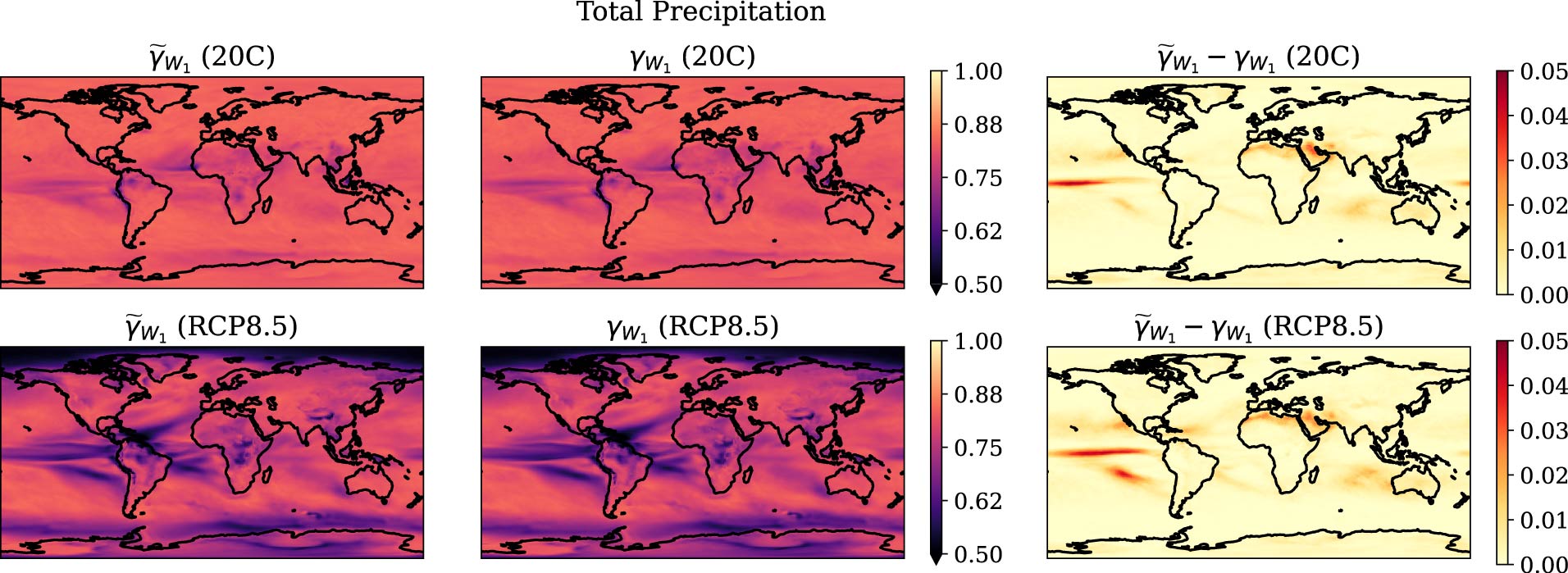}
    \caption{
        As in Fig.~\ref{figS01}, but for the total precipitation. The middle column reproduces the $\gamma_{W_1}$ panels of Fig.~\ref{fig05} for comparison.
    }
    \label{figS02}
\end{figure*}

The difference between $\gamma_{W_1}$ and $\widetilde{\gamma}_{W_1}$ is small in practice.
In every synthetic experiment, with and without outliers, $\widetilde{\gamma}_{W_1}-\gamma_{W_1}$ remains below 0.02 (not shown).
For CESM-LE, Figs.~\ref{figS01} and~\ref{figS02} compare the two metrics.
Their maps are nearly identical for both the 2~m air temperature and the total precipitation, under the 20C and the RCP8.5 forcing.
The difference $\widetilde{\gamma}_{W_1}-\gamma_{W_1}$ is positive everywhere, consistent with the ordering above, and stays below 0.05.
Its largest values appear in the eastern equatorial Pacific for the total precipitation, a variable whose distribution is heavy-tailed \citep{Franzke+20}.
Additivity of the two transport costs is therefore not exact, but the departure from it is small enough that the two metrics $\gamma_{W_1}$ and $\widetilde{\gamma}_{W_1}$ support the same conclusions.

%TC:ignore
%TC:endignore

%TC:ignore
%TC:endignore

\FloatBarrier

\subsection*{S3 Generation of the synthetic climate data} \label{subsec:synthetic-data-generation}

The synthetic climate data \citep{Sane+24} are independent and identically distributed in time, and the underlying distribution is Gaussian, uniform, or lognormal.
The Gaussian distribution has zero mean and unit variance, the uniform distribution has support $[-1,1]$, and the lognormal distribution has zero mean and unit variance in its logarithm.
For each distribution, correlation between ensemble members was imposed by setting the Pearson correlation coefficient $\rho$ to a fixed value in $[0,1)$.
If the time series is sufficiently long, the correlation coefficient between any pair of members is approximately $\rho$.
We varied $\rho$ in steps of 0.001 from 0 to 0.009, in steps of 0.01 from 0.01 to 0.99, and in steps of 0.001 from 0.99 to 0.999.
We describe below how the correlated random numbers were generated.

The Gaussian random numbers were generated by setting all off-diagonal elements of the covariance matrix to $\rho$.
The covariance matrix is $N_{\mathrm{ens}} \times N_{\mathrm{ens}}$, and its diagonal elements (i.e., the variances) are all unity.

The uniform random numbers were generated using a Gaussian copula \citep[e.g.,][]{Nelsen06}.
Gaussian random numbers with a correlation coefficient of $\rho$ were first generated as above.
Applying the cumulative distribution function of the standard Gaussian, we transformed them into sequences following the standard uniform distribution.
A linear transformation was then applied to change the support from $[0,1]$ to $[-1,1]$.
Due to the nonlinearity of the cumulative distribution function, the resulting uniform random numbers have a correlation coefficient of only approximately $\rho$.

The lognormal random numbers were generated using the relation between the lognormal and Gaussian distributions.
If two standard lognormal variables have a correlation coefficient of $\rho$, then the Gaussian variables obtained by the logarithmic transformation have a correlation coefficient of $\rho'$ \citep[e.g.,][]{Crow+88}:
\begin{align*}
    \rho' = \ln \left[ 1 + \rho \left( e - 1 \right) \right].
\end{align*}
On the basis of this relation, we generated Gaussian random numbers with $\rho'$ as above and applied the exponential transformation to them, thereby obtaining lognormal random numbers with a correlation coefficient of $\rho$.

\subsection*{S4 Synthetic climate data with temporal correlation}

To examine the effect of temporal correlation, we also generated time series following a first-order autoregressive [AR(1)] process.
The Gaussian random numbers $\varepsilon_{n,t}$ generated as described in Section~S3 were converted into
\begin{align*}
    z_{n,1} = \varepsilon_{n,1}, \qquad
    z_{n,t} = \phi\, z_{n,t-1} + \sqrt{1-\phi^2}\, \varepsilon_{n,t} \quad (t = 2,\dots,N_T),
\end{align*}
where $\phi \in [0,1)$ is the lag-one autocorrelation of $z_{n,t}$ at the monthly time step.
For the uniform and lognormal distributions, the resulting $z_{n,t}$ were then transformed as explained above.
This procedure keeps the distribution at each time step and the instantaneous correlation between members unchanged, while introducing temporal correlation into each member's time series.

\begin{figure*}[tb]
    \centering
    \includegraphics[width=1.0\textwidth]{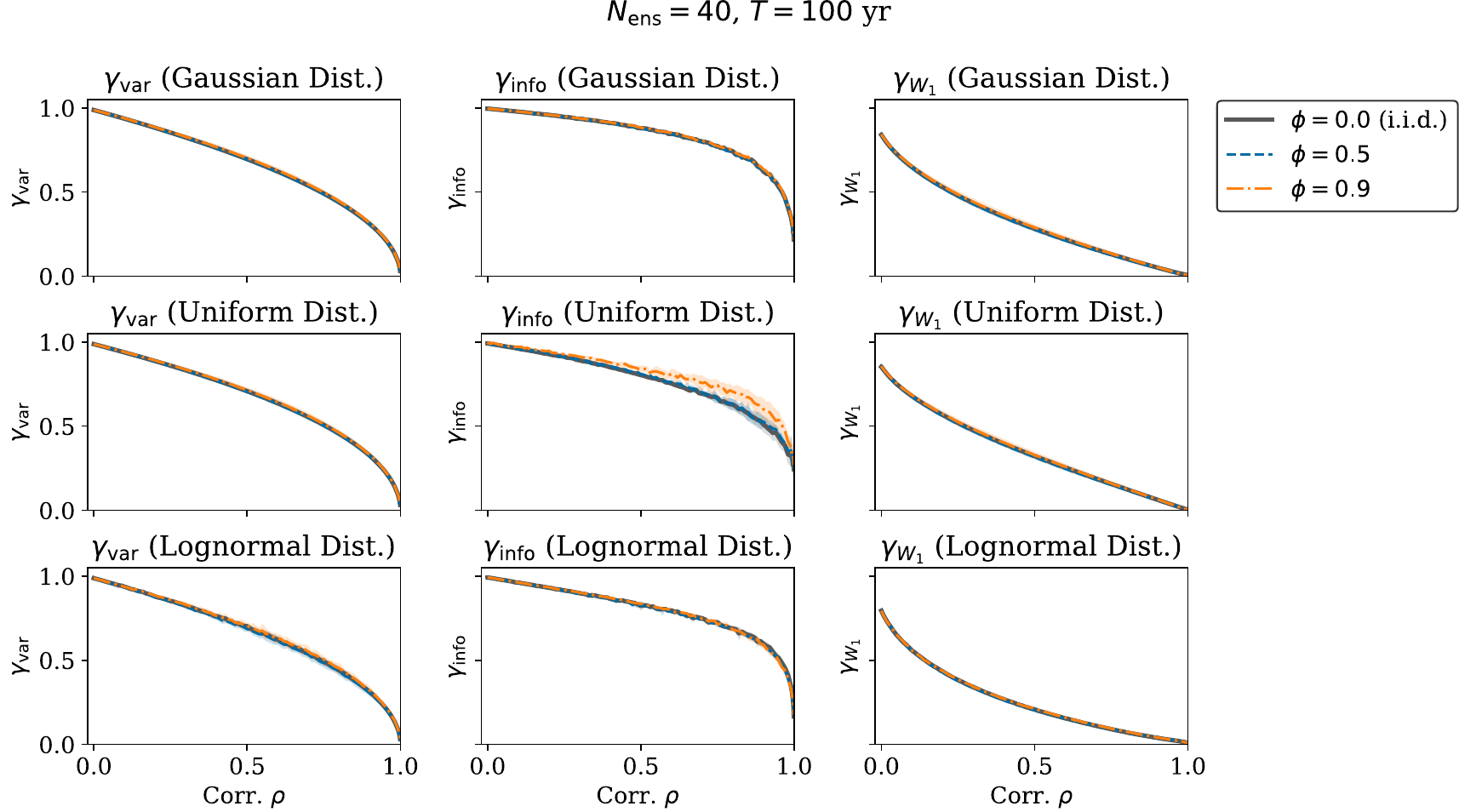}
    \caption{
        Dependence of the three metrics on the temporal correlation of the synthetic climate data.
        The results are for $N_{\mathrm{ens}}=40$ and $T=100$~years without outliers.
        Columns show the metrics ($\gamma_{\mathrm{var}}$, $\gamma_{\mathrm{info}}$, and $\gamma_{W_1}$, from left to right) and rows the Gaussian, uniform, and lognormal distributions.
        Solid, dashed, and dash-dotted lines show the lag-one autocorrelation $\phi = 0$ (no temporal correlation), 0.5, and 0.9, respectively.
        The temporal correlation increases with $\phi$.
        Each line is the mean over five random seeds, and shading indicates one standard deviation above and below the mean.
    }
    \label{figS03}
\end{figure*}

Figure~\ref{figS03} shows the results for $\phi = 0$, 0.5, and 0.9 with $N_{\mathrm{ens}}=40$ and $T=100$~years, where $\phi = 0$ (i.e., no temporal correlation) corresponds to the case examined in the main text.
For all metrics and distributions, the curves for the three values of $\phi$ nearly overlap.
The exception is $\gamma_{\mathrm{info}}$ for the uniform distribution at $\phi=0.9$, which departs from the other curves at large $\rho$.
This departure is likely because the smaller effective sample size makes the estimated bin width fluctuate more.

%TC:ignore
%TC:endignore

\FloatBarrier

\subsection*{S5 Additional results for the synthetic climate data}

\begin{figure*}[tb]
    \centering
    \includegraphics[width=1.0\textwidth]{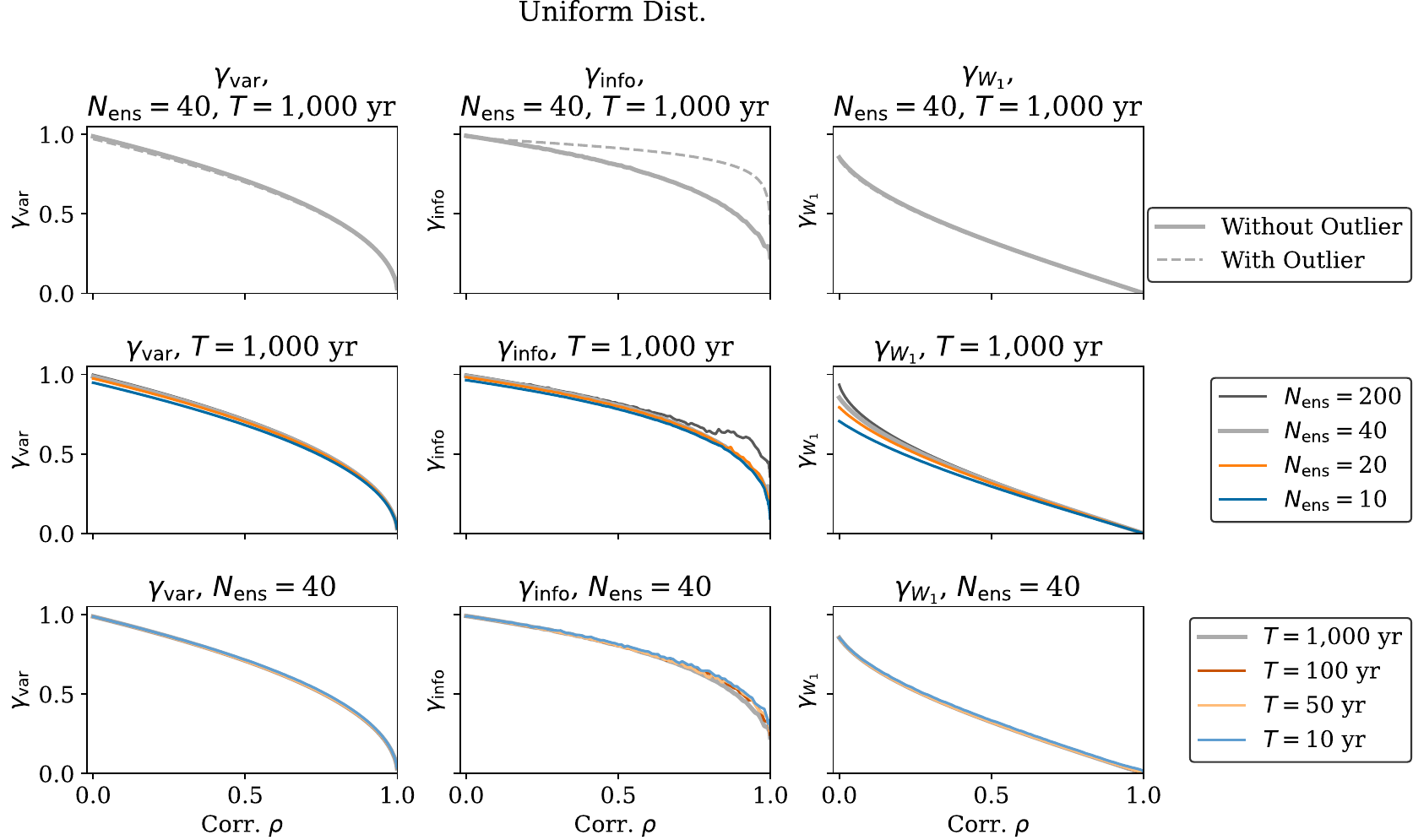}
    \caption{
        Sensitivity of the three metrics for synthetic data based on the uniform distribution.
        Each column shows one metric ($\gamma_{\mathrm{var}}$, $\gamma_{\mathrm{info}}$, and $\gamma_{W_1}$, from left to right) as a function of the correlation coefficient $\rho$.
        Top: with (dashed) and without (solid) outliers, for $N_{\mathrm{ens}}=40$ and $T=1{,}000$~years.
        Middle and bottom: four values of the ensemble size $N_{\mathrm{ens}}$ ($T=1{,}000$~years) and of the data (time series) length $T$ ($N_{\mathrm{ens}}=40$), respectively, both without outliers.
    }
    \label{figS04}
\end{figure*}

\begin{figure*}[tb]
    \centering
    \includegraphics[width=1.0\textwidth]{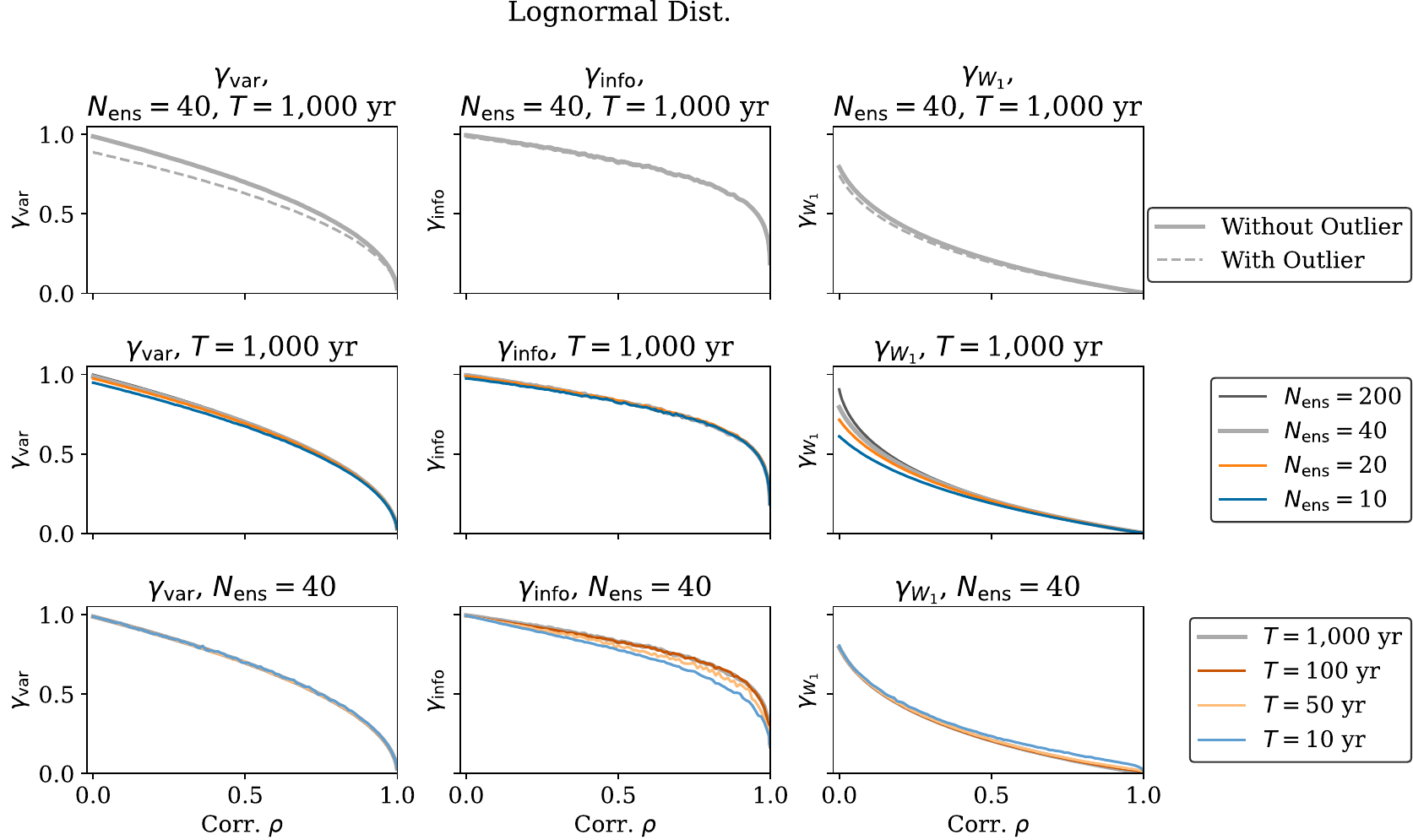}
    \caption{
        As in Fig.~\ref{figS04}, but for synthetic data based on the lognormal distribution.
    }
    \label{figS05}
\end{figure*}

Figures~\ref{figS04} and~\ref{figS05} show the results for the synthetic climate data based on the uniform and lognormal distributions, respectively.
Note that Fig.~\ref{fig02} in the main text shows the results based on the Gaussian distribution.
The panel layout is as in Fig.~\ref{fig02}.
The results depend on the shape of the probability distribution.
$\gamma_{\mathrm{var}}$ is in some cases sensitive to outliers while remaining nearly independent of $T$.
The metric $\gamma_{\mathrm{info}}$ is sensitive to $T$, and in some cases to $N_{\mathrm{ens}}$ and outliers as well.
Outliers affect $\gamma_{\mathrm{var}}$ except for the uniform distribution, and $\gamma_{\mathrm{info}}$ except for the lognormal.
In contrast, $\gamma_{W_1}$ is nearly independent of outliers and $T$, although it is sensitive to $N_{\mathrm{ens}}$ for $N_{\mathrm{ens}} \lesssim 20$.
These features of $\gamma_{W_1}$ are common to all three distributions.

%TC:ignore
%TC:endignore

%TC:ignore
%TC:endignore

\begin{figure*}[tb]
    \centering
    \includegraphics[width=1.0\textwidth]{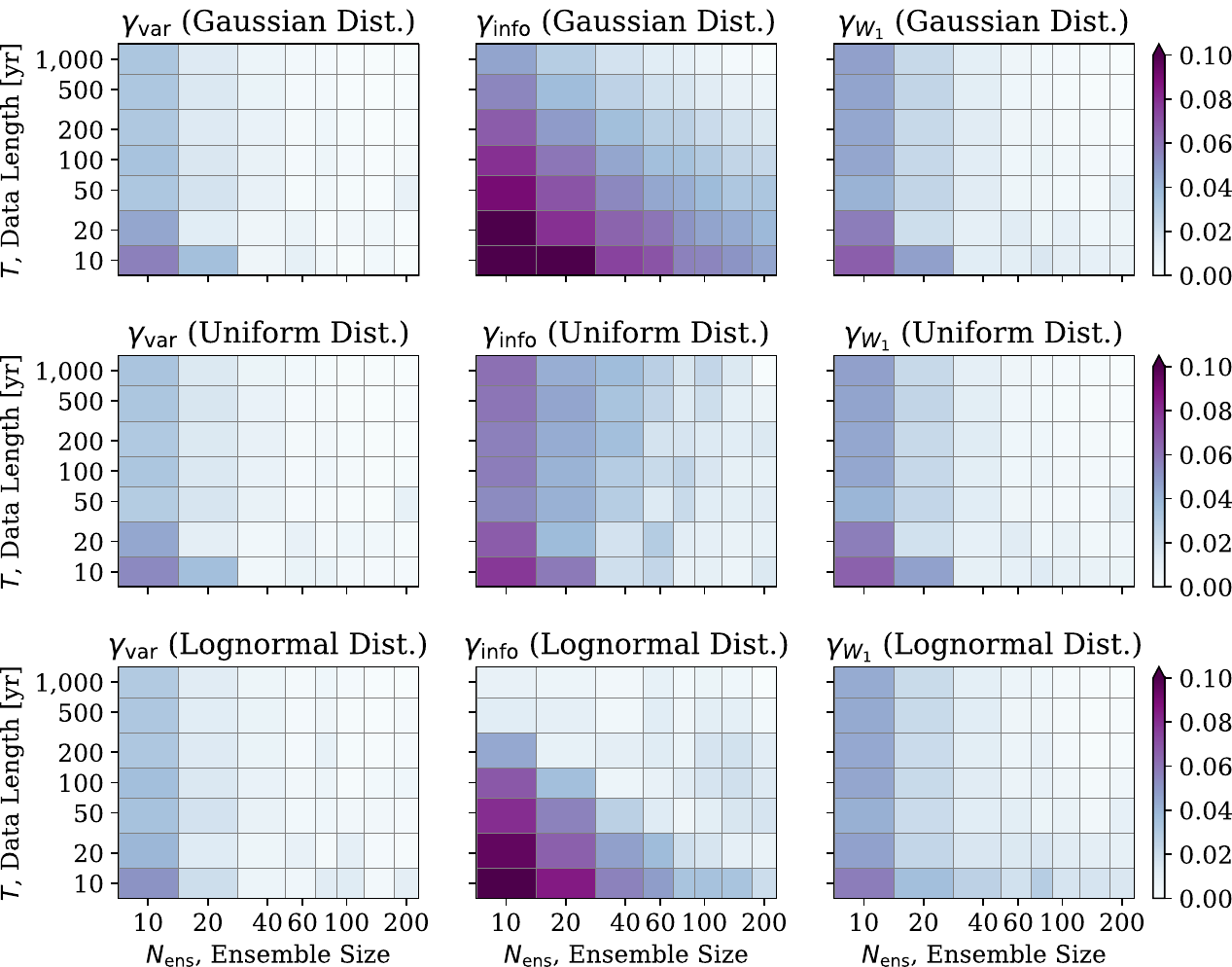}
    \caption{
        Departure of the three metrics from the reference for synthetic data without outliers.
        Columns show the metrics ($\gamma_{\mathrm{var}}$, $\gamma_{\mathrm{info}}$, and $\gamma_{W_1}$, from left to right) and rows the Gaussian, uniform, and lognormal distributions.
        Each panel gives the departure on a grid of the ensemble size $N_{\mathrm{ens}}$ and the data (time series) length $T$.
        The reference is the value obtained for $N_{\mathrm{ens}}=200$ and $T=1{,}000$~years, and the departure is the absolute difference from it integrated over $\rho \in [0,1)$.
    }
    \label{figS06}
\end{figure*}

The departure from the reference is defined as in ``\nameref{sec:results}.''
Figure~\ref{figS06} shows the dependence on $N_{\mathrm{ens}}$ and $T$ without outliers.
Note that Fig.~\ref{fig03} in the main text shows the case with outliers.
Without outliers (Fig.~\ref{figS06}), $\gamma_{\mathrm{var}}$ is the least sensitive of the three metrics to $N_{\mathrm{ens}}$ and $T$, and $\gamma_{\mathrm{info}}$ is the most sensitive.
As in Fig.~\ref{fig03}, the departure of the proposed metric $\gamma_{W_1}$ from the reference depends only weakly on the distribution shape, and $\gamma_{W_1}$ gives stable estimates for $N_{\mathrm{ens}} \sim 40$ or more.

%TC:ignore
%TC:endignore

\FloatBarrier

\subsection*{S6 Convergence of the reference}

\begin{figure*}[tb]
    \centering
    \includegraphics[width=0.75\textwidth]{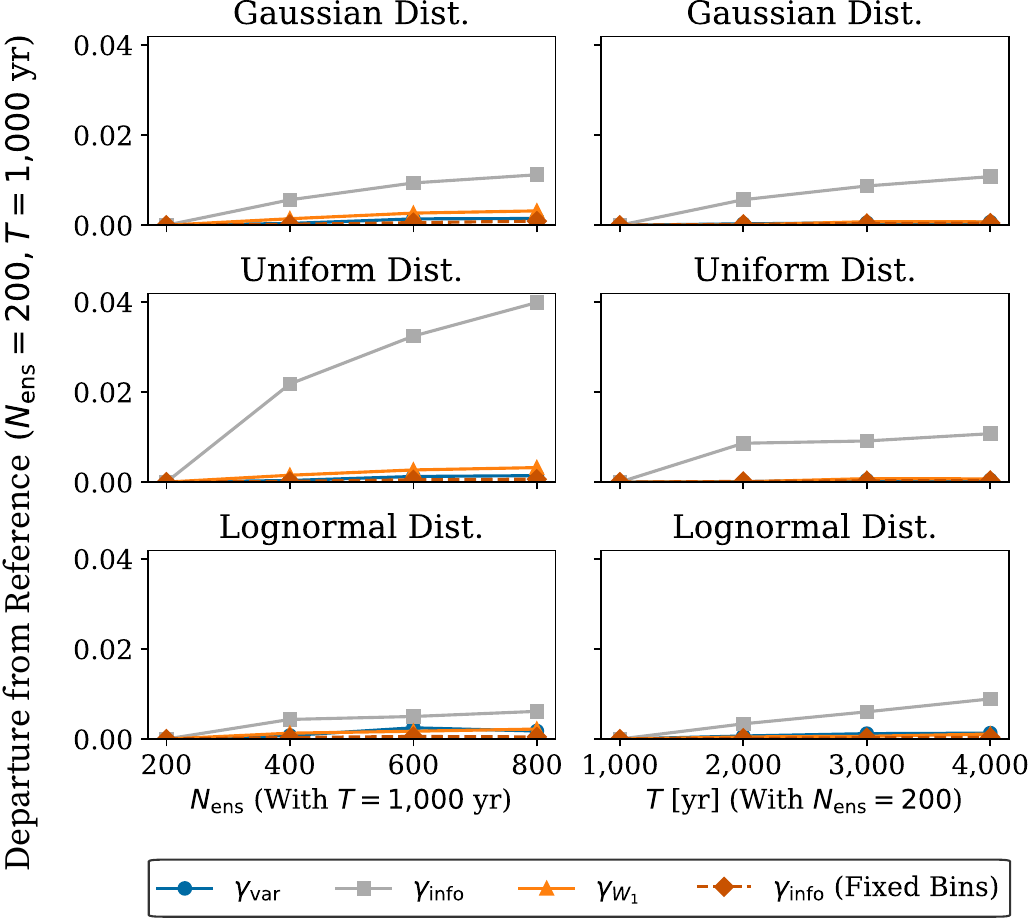}
    \caption{
        Convergence of the reference for synthetic data without outliers.
        The reference is the value obtained for $N_{\mathrm{ens}}=200$ and $T=1{,}000$~years.
        The left column shows the departure from the reference for increasing $N_{\mathrm{ens}}$ with $T=1{,}000$~years, and the right column for increasing $T$ with $N_{\mathrm{ens}}=200$.
        Rows show the Gaussian, uniform, and lognormal distributions.
        The departure is the absolute difference from the reference integrated over $\rho \in [0,1)$.
        Dashed lines show $\gamma_{\mathrm{info}}$ with the number of bins and the range fixed.
        All values are computed from the means over five random seeds.
    }
    \label{figS07}
\end{figure*}

To examine whether the reference ($N_{\mathrm{ens}}=200$ and $T=1{,}000$~years) has converged, we increased either $N_{\mathrm{ens}}$ to 400, 600, and 800 with $T=1{,}000$~years or $T$ to 2,000, 3,000, and 4,000~years with $N_{\mathrm{ens}}=200$, without outliers.
Figure~\ref{figS07} shows the departure from the reference, defined as in ``\nameref{sec:results}.''
For $\gamma_{\mathrm{var}}$ and $\gamma_{W_1}$, the departure remains small for all three distributions, which indicates that the reference has sufficiently converged.
For $\gamma_{\mathrm{info}}$, the departure grows with $N_{\mathrm{ens}}$ and $T$, because the bin width estimated from the data changes with the sample size.
When the number of bins and the range are fixed, the departure of $\gamma_{\mathrm{info}}$ also remains small (dashed lines).
The fixed binning follows the source code of \cite{Sane+25} for the Gaussian and uniform distributions, with the ranges $[-5,5]$ and $[-1,1]$, respectively.
For the lognormal distribution, the range $[0,20]$ was chosen to include the outlier value used in the main experiments.
To compute $\gamma_{\mathrm{info}}$ with this fixed binning, we used 60 bins in all cases, following \cite{Sane+25}.

%TC:ignore
%TC:endignore

\FloatBarrier

\subsection*{S7 Dependence of $\gamma_{\mathrm{info}}$ on the bin width}

The metric $\gamma_{\mathrm{info}} = 1 - I(X;M)/H(X)$ [Eq.~(\ref{eq:gamma_info})] is computed from histograms with bins of width $\Delta$.
As $\Delta$ decreases, the mutual information $I(X;M)$ converges to the corresponding value for the continuous variables, whereas the Shannon entropy $H(X)$ keeps increasing \citep[e.g.,][]{Cover+Thomas05}.
Specifically, for narrow bins,
\begin{align*}
    H(X) \approx h(X) - \log_2 \Delta,
\end{align*}
where $h(X) = -\int [p(x) \log_2 p(x)]\,\mathrm{d}x$ is the differential entropy of $X$, and $p(x)$ is the probability density of $X$.
Substituting this into Eq.~(\ref{eq:gamma_info}) gives
\begin{align*}
    \gamma_{\mathrm{info}} = 1-\frac{I(X;M)}{H(X)} \approx 1-\frac{I(X;M)}{h(X) - \log_2 \Delta}.
\end{align*}
In this ratio, the numerator $I(X;M)$ does not depend on $\Delta$, whereas the denominator increases as $\Delta$ decreases, which suggests that $\gamma_{\mathrm{info}}$ reaches 1.
In the synthetic data, the distribution of $X$, and hence $h(X)$, does not change with $\rho$.
The change of $\gamma_{\mathrm{info}}$ with $\rho$ thus comes only from $I(X;M)$ and becomes smaller if the estimated $\Delta$ decreases.

\subsection*{S8 Dependence on the fraction of outliers}

\begin{figure*}[tb]
    \centering
    \includegraphics[width=1.0\textwidth]{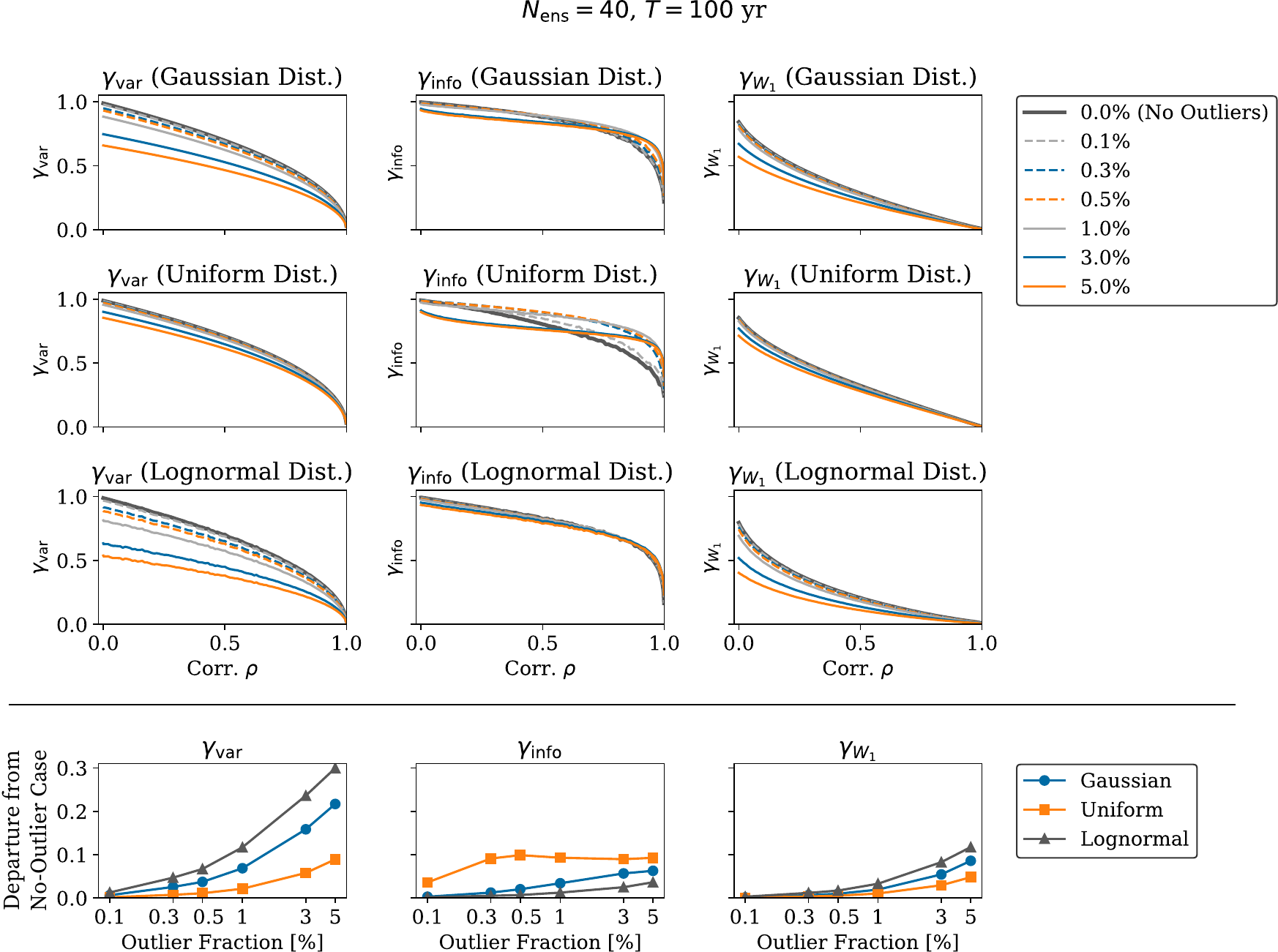}
    \caption{
        Dependence of the three metrics on the fraction of outliers for synthetic data with $N_{\mathrm{ens}}=40$ and $T=100$~years.
        In the top three rows, columns show the metrics ($\gamma_{\mathrm{var}}$, $\gamma_{\mathrm{info}}$, and $\gamma_{W_1}$, from left to right) and rows the Gaussian, uniform, and lognormal distributions.
        Each of these panels shows the metric as a function of $\rho$ for outlier fractions from 0\% (no outliers) to 5\%.
        The bottom row shows the departure of each metric from the case without outliers as a function of the outlier fraction.
        The departure is the absolute difference integrated over $\rho \in [0,1)$.
        All values are computed from the means over five random seeds.
    }
    \label{figS08}
\end{figure*}

To examine the dependence on the fraction of outliers, we varied it over 0.1, 0.3, 0.5, 1, 3, and 5\% of all samples for $N_{\mathrm{ens}}=40$ and $T=100$~years (Fig.~\ref{figS08}), where 0.5\% is the value used in the main text.
Here, the departure is measured from the case without outliers for the same $N_{\mathrm{ens}}$ and $T$.
For all three distributions, the departure of $\gamma_{\mathrm{var}}$ and $\gamma_{W_1}$ increases gradually and monotonically with the fraction, and that of $\gamma_{W_1}$ is about one-quarter to one-half of that of $\gamma_{\mathrm{var}}$.
In contrast, the departure of $\gamma_{\mathrm{info}}$ can be large even for the smallest fraction and does not always increase with the fraction.
Indeed, for the uniform distribution, the departure of $\gamma_{\mathrm{info}}$ is already 0.036 at 0.1\% and 0.091 at 0.3\%, almost the same as at 5\% (0.093), whereas that of $\gamma_{W_1}$ is 0.0035 at 0.3\% and 0.048 at 5\%.
This is likely because the outliers change the bin width estimated from the data.
For the lognormal distribution, however, $\gamma_{\mathrm{info}}$ is the least affected of the three metrics.
The results for $N_{\mathrm{ens}}=40$ with $T=1{,}000$~years and for $N_{\mathrm{ens}}=200$ with $T=100$~years are similar for $\gamma_{\mathrm{var}}$ and $\gamma_{W_1}$, whereas those for $\gamma_{\mathrm{info}}$ change with the sample size (not shown).

%TC:ignore
%TC:endignore

\FloatBarrier

\subsection*{S9 Slope of the metrics with respect to $\rho$}

\begin{figure*}[tb]
    \centering
    \includegraphics[width=1.0\textwidth]{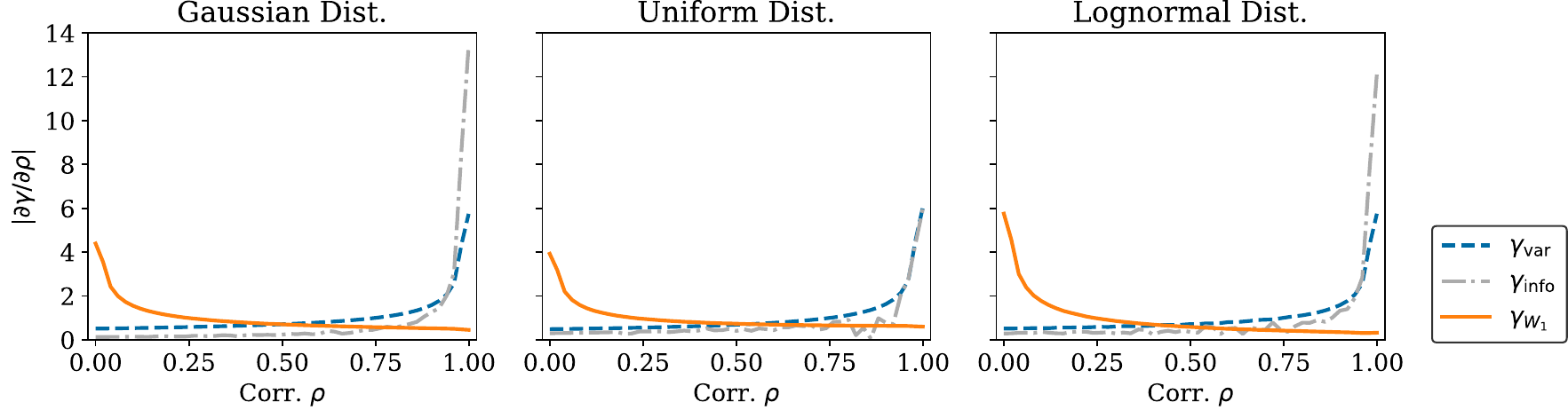}
    \caption{
        Absolute slope of the three metrics with respect to the correlation coefficient $\rho$.
        Panels show the Gaussian, uniform, and lognormal distributions, from left to right.
        Dashed, dash-dotted, and solid lines show $\gamma_{\mathrm{var}}$, $\gamma_{\mathrm{info}}$, and $\gamma_{W_1}$, respectively.
        The slope $\lvert \partial\gamma/\partial\rho \rvert$ is computed by central differences at intervals of 0.02 in $\rho$ from the curves averaged over five random seeds, for synthetic data with $N_{\mathrm{ens}}=200$ and $T=1{,}000$~years without outliers.
    }
    \label{figS09}
\end{figure*}

To evaluate how rapidly each metric changes with $\rho$, we computed the absolute slope $\lvert \partial\gamma/\partial\rho \rvert$ of the three metrics (Fig.~\ref{figS09}) for the reference setting ($N_{\mathrm{ens}}=200$ and $T=1{,}000$~years, without outliers).
The slope was obtained by central differences (one-sided at the end points) of the curves averaged over five random seeds, at intervals of 0.02 in $\rho$ from 0 to 0.98 and at $\rho = 0.999$.
For all three distributions, the slope of $\gamma_{W_1}$ is largest near $\rho = 0$ and decreases as $\rho$ increases, whereas that of $\gamma_{\mathrm{var}}$ increases toward $\rho = 1$.
The slopes of these two metrics cross near $\rho = 0.5$.
For the Gaussian distribution, the analytic forms of the two metrics give the same crossing at $\rho = 0.5$ (Section~S10).
The slope of $\gamma_{\mathrm{info}}$ is the smallest of the three for $\rho \le 0.5$ and increases sharply near $\rho = 1$.

%TC:ignore
%TC:endignore

\FloatBarrier

\subsection*{S10 Analytic forms of $\gamma_{\mathrm{var}}$ and $\gamma_{W_1}$ for the Gaussian distribution} \label{subsec:gaussian-analytic}

For the synthetic climate data based on the Gaussian distribution, $\gamma_{\mathrm{var}}$ and $\gamma_{W_1}$ can be written as functions of $\rho$ in the limit of $T \to \infty$ and $N_{\mathrm{ens}} \to \infty$.
At each time step, $x_{n,t}$ follows the Gaussian distribution with zero mean and unit variance, and the correlation coefficient between any pair of members is $\rho$ (Section~S3).
Since $m_t$ is the ensemble mean of $x_{n,t}$, it also follows the Gaussian distribution with zero mean, and its variance is
\begin{align*}
    \mathrm{Var}(m_t) = \frac{1}{N_{\mathrm{ens}}^2}\sum_{n=1}^{N_{\mathrm{ens}}}\sum_{n'=1}^{N_{\mathrm{ens}}}\mathrm{Cov}(x_{n,t},x_{n',t}) = \frac{N_{\mathrm{ens}} + N_{\mathrm{ens}}(N_{\mathrm{ens}}-1)\rho}{N_{\mathrm{ens}}^2} \to \rho \quad (N_{\mathrm{ens}} \to \infty).
\end{align*}
The total mean $\overline{m}$ (i.e., the temporal mean of $m_t$) is zero in this limit.

The metric $\gamma_{\mathrm{var}}$ is computed from $\sigma_{\mathrm{int}}^2$ and $\sigma_{\mathrm{ext}}^2$.
Here, $\sigma_{\mathrm{ext}}^2$ (i.e., the variance of $m_t$) equals $\rho$, and $\sigma_{\mathrm{int}}^2 + \sigma_{\mathrm{ext}}^2$ (i.e., the variance of $x_{n,t}$) equals unity, leading to $\sigma_{\mathrm{int}}^2 = 1 - \rho$.
Therefore, $\gamma_{\mathrm{var}}$ is obtained as follows:
\begin{align*}
    \gamma_{\mathrm{var}} &= \frac{\sqrt{\sigma_{\mathrm{int}}^2}}{\sqrt{\sigma_{\mathrm{int}}^2+\sigma_{\mathrm{ext}}^2}} = \sqrt{1-\rho}.
\end{align*}

For $\gamma_{W_1}$, the distribution of $m_t$ is that of $x_{n,t}$ contracted toward zero by the factor $\sqrt{\rho}$ because both $m_t$ and $x_{n,t}$ are Gaussian with zero mean and their standard deviations are $\sqrt{\rho}$ and 1, respectively.
This means that the quantiles satisfy $m_{(i)} = \sqrt{\rho}\,x_{(i)}$.
Substituting these quantiles into Eqs.~(\ref{eq:w_int}) and~(\ref{eq:w_ext}) gives $w_{\mathrm{int}} = (1-\sqrt{\rho})\,w_{\mathrm{tot}}$ and $w_{\mathrm{ext}} = \sqrt{\rho}\,w_{\mathrm{tot}}$, where $w_{\mathrm{tot}}$ is the direct transport cost defined above and equals the average of $\lvert x_{(i)} \rvert$ over all samples.
Therefore, $\gamma_{W_1}$ is obtained as follows:
\begin{align*}
    \gamma_{W_1} = \frac{w_{\mathrm{int}}}{w_{\mathrm{int}}+w_{\mathrm{ext}}} = 1-\sqrt{\rho}.
\end{align*}

The absolute slopes of the two metrics with respect to $\rho$ are
\begin{align*}
    \left\lvert \frac{\partial\gamma_{\mathrm{var}}}{\partial\rho} \right\rvert &= \frac{1}{2\sqrt{1-\rho}}, \\
    \left\lvert \frac{\partial\gamma_{W_1}}{\partial\rho} \right\rvert &= \frac{1}{2\sqrt{\rho}}.
\end{align*}
The slope of $\gamma_{W_1}$ is larger for $\rho < 0.5$, and that of $\gamma_{\mathrm{var}}$ for $\rho > 0.5$, the two being equal at $\rho = 0.5$.
This result is consistent with the numerical slopes for $N_{\mathrm{ens}} = 200$ and $T = 1{,}000$~years (Fig.~\ref{figS09}).

\end{appendices}

%TC:ignore
\bibliography{references}
%TC:endignore

\end{document}